\documentclass[11pt]{article}
\usepackage[T1]{fontenc}
\usepackage[utf8]{inputenc}
\usepackage{mathtools} % loads amsmath; gives bmatrix* etc.
\usepackage{amssymb}
\usepackage{graphicx,psfrag,epsf}
\usepackage{caption}
\usepackage{subcaption}
\usepackage{enumerate}
\usepackage{natbib}
\usepackage{dcolumn}

\usepackage{color}
\usepackage{url}
\newcommand{\blind}{0}

\usepackage{geometry}
\numberwithin{equation}{section}
\numberwithin{figure}{section}
\numberwithin{table}{section}

\DeclareMathOperator\erf{erf}

\newcommand{\dd}{\textrm{d}}
\newcommand{\Cpp}{C\nolinebreak\hspace{-.05em}\raisebox{.4ex}{\tiny\bf +}\nolinebreak\hspace{-.10em}\raisebox{.4ex}{\tiny\bf +}}

\newcolumntype{d}[1]{D{.}{.}{#1}} \newcolumntype{t}[1]{D{-}{\times}{#1}}

\newcommand{\MVN}{\mathop{\mathcal{N}}}
\newcommand{\Norm}{\mathop{\mathcal{N}}}
\newcommand{\diag}{\mathop{\mathrm{diag}}}
\newcommand{\chivec}{\boldsymbol{\chi}}
\newcommand{\yvec}{\boldsymbol{y}}
\newcommand{\mupop}{\boldsymbol{\mu}}
\newcommand{\covpop}{C_{\rm pop}}

\newcommand{\assm}{\,|\,}
\newcommand{\mlike}{\ell}
\newcommand{\ppar}{\theta}  % pop'n parameters
\newcommand{\obsv}{\mathcal{O}}  % observables
\newcommand{\psivec}{\boldsymbol{\psi}}

\newcommand{\lfunc}{\phi}  % or \Phi or \varphi or...

\newcommand{\lpdf}{f}
\newcommand{\lcdf}{F}
\newcommand{\rhopar}{\boldsymbol{\zeta}}

\newcommand{\Fth}{F_{\rm th}}
\newcommand{\Fhat}{\hat{F}}
\newcommand{\Ffid}{F_{\rm fid}}
\newcommand{\rmax}{r_{\rm max}}

\newcommand{\aeffic}{\overline{\eta}}
\newcommand{\like}{\mathcal{L}}
\newcommand{\epdf}{\mu}
\newcommand{\Lsol}{L_\odot}

\newcommand\enote[1]{{$\bullet\bullet\bullet$}{\sl [#1]}{$\bullet\bullet\bullet$}}
\begin{document}

\def\spacingset#1{\renewcommand{\baselinestretch}%
{#1}\small\normalsize} \spacingset{1}

\if0\blind
{
  %\title{\bf CUDAHM:  GPU-Accelerated Bayesian Inference for Simple Hierarchical Models}
  %\title{\bf CUDAHM:  GPU-Accelerated Hierarchical Bayesian Modeling of Cosmic Populations}
  \title{\bf GPU-Accelerated Hierarchical Bayesian Inference
  with Application to Modeling Cosmic Populations: CUDAHM}
  \author{J\'anos M. Szalai-Gindl\\
    %\thanks{
    %\textit{This work was supported by the Hungarian Scientific Research Fund via grant OTKA~NN~114560.  Budav\'ari, Kelly, and Loredo gratefully acknowledge the NSF-funded Statistical and Applied Mathematical Sciences Institute (SAMSI) for support for visits to SAMSI, where this project originated. Loredo's effort was additionally supported by NSF grant AST-1312903.}}\hspace{.2cm}\\
    Department of Information Systems,\\
    E\"otv\"os Lor\'and University\\
    and \\
    Thomas J. Loredo \\
    Cornell Center for Astrophysics \& Planetary Science,\\
    Cornell University \\
    and \\
    Brandon C. Kelly \\
    Department of Physics,\\
    University of California, Santa Barbara\\  % Broida Hall, 
    and \\
    Istv\'an Csabai\\
    Department of Physics of Complex Systems,\\
    E\"otv\"os Lor\'and University\\
    and \\
    Tam\'as Budav\'ari \\
    Department of Applied Mathematics \& Statistics,\\
    The Johns Hopkins University\\
    and \\
    L\'aszl\'o Dobos\\
    Department of Physics of Complex Systems,\\
    E\"otv\"os Lor\'and University
}
  \date{July 31, 2018}   
  \maketitle
} \fi

\if1\blind
{
  \bigskip
  \bigskip
  \bigskip
  \begin{center}
    {\LARGE\bf CUDAHM:  GPU-Accelerated Hierarchical Bayesian Inference}
\end{center}
  \medskip
} \fi

\bigskip
\begin{abstract}
We describe a computational framework for hierarchical Bayesian inference with simple (typically single-plate) parametric graphical models that uses graphics processing units (GPUs) to accelerate computations, enabling deployment on very large datasets.
Its \Cpp\ implementation, CUDAHM (CUDA for Hierarchical Models) exploits conditional independence between instances of a plate, facilitating massively parallel exploration of the replication parameter space using the single instruction, multiple data architecture of GPUs.
It provides support for constructing Metropolis-within-Gibbs samplers that iterate between GPU-accelerated robust adaptive Metropolis sampling of plate-level parameters conditional on upper-level parameters, and Metropolis-Hastings sampling of upper-level parameters on the host processor conditional on the GPU results.
CUDAHM is motivated by demographic problems in astronomy, where density estimation and linear and nonlinear regression problems must be addressed for populations of thousands to millions of objects whose features are measured with possibly complex uncertainties.
We describe a thinned latent point process framework for modeling such demographic data.
We demonstrate accurate GPU-accelerated parametric conditional density deconvolution for simulated populations of up to $3\times10^5$ objects in $\approx1$ hour using a single NVIDIA Tesla K40c GPU.
Supplementary material provides details about the CUDAHM API and the demonstration problem.
\end{abstract}

% The GPU computations are implemented using the Compute Unified Device Architecture (CUDA) parallel computing platform.

\noindent%

%\vfill
%\newpage

% *** Comment this for arXiv version, for single-space
%\spacingset{1.45} % DON'T change the spacing!

% Just for posting to arXiv; also add appendices:
% \spacingset{1}

%===============================================================================
\section{Introduction}
\label{sec:intro}

Bayesian inference with graphical models has rapidly grown in popularity and sophistication since the emergence of Markov chain Monte Carlo (MCMC) algorithms for Bayesian computation nearly three decades ago.
The work we report here focuses on models with classic, simple graphical structures---directed acyclic graphs (DAGs) that typically have a single plate, i.e., a single level of replication of random variables at the lower level of a hierarchical model.
Our work aims to extend the range of application of Bayesian graphical modeling in the direction of increased dataset size, rather than in the direction of increased graphical complexity.

We are motivated by univariate and multivariate measurement error problems in astronomy: density estimation with measurement error (density deconvolution, or demixing, often of a \emph{conditional} density), and linear and nonlinear regression with measurement errors in both predictors and response.
Hierarchical Bayesian modeling is well-suited to such problems, but is relatively new in astronomy (see \citealt{kelly-measurement2012} and \citealt{loredo2013survey} for recent surveys).
Although some recent astrostatistical research develops models with rich graphical structure, most astronomers are unfamiliar with hierarchical modeling, and models with simple graphical structure can provide new capability in many areas of astronomy, including basic two-level hierarchical models.%
\footnote{We follow the convention of naming hierarchical DAGs by the number of levels with uncertain nodes, e.g., the number of open nodes in Fig.~\ref{fig:DAG-2Level}, which depicts a model with three levels of random variables, but with the data variables (bottom level) known, i.e., to be conditioned on.}
But dataset size can be an obstacle to use of such models.
Large-scale, automated astronomical surveys are providing astronomers with increasingly large datasets for demographic studies of cosmic populations (e.g., categories of stars, galaxies, planets, and minor planets).
Current and emerging surveys are providing measurements for populations with sizes ranging from tens of thousands to $10^8$ or even larger.
For datasets of these scales, exploration or integration over the latent variables specifying imprecisely measured characteristics of objects in a population can be prohibitive, even for simple models with univariate member characteristics.
Yet as population size grows, it becomes increasingly important to account for uncertainty in such latent variables.
For example, it is well known that regression and density estimators that ignore measurement error are typically inconsistent, with the ratio of bias to reported precision growing with sample size \citep{C+06-MsmtErr}.
Single-plate hierarchical models can account for measurement error in many astronomically interesting scenarios, provided the implementation enables efficient computation for relevant dataset sizes.

%Section~\ref{sec:app_models_methods} describes the applied models and methods in general.

In the following section we describe the design of a \Cpp\ framework, CUDAHM (abbreviating CUDA%
\footnote{CUDA signifies the Compute Unified Device Architecture (CUDA) parallel computing platform for NVIDIA GPUs.
It comprises the CUDA language extensions for C and \Cpp, as well as a compiler and other tools for development of general-purpose GPU algorithms running on NVIDIA GPUs.
It includes several GPU-optimized scientific computing libraries; CUDAHM uses CUDA's linear algebra and pseudo-random number generator libraries.}
framework for Hierarchical Modeling), which is motivated by a prevalent computational structure underlying example hierarchical models arising in measurement error problems in astronomy. 
%\enote{Perhaps add a brief section (\S~3) treating Brandon's nonlinear regression with measurement error problem; perhaps he'd be willing to write it; a very brief summary of it is in \S~2 and may suffice, with Brandon's example presented in the followup A\&C paper.}
Sec.~\ref{sec:perform} describes results of a simple performance test, applying CUDAHM to the normal-normal hierarchical model, and comparing performance to a CPU-based implementation using the Stan probabilistic computing language.
In Sec.~\ref{sec:lum_func}, we describe a common astronomical data analysis problem: inferring the luminosity distribution of a class of objects from distance and flux observations of a sample subject to both measurement error and selection effects.
We present a \emph{thinned latent marked point processes} framework for such problems, with latent parameters accounting for measurement error, and thinning accounting for selection.
We show that the resulting likelihood function can be cast in a form mirroring the computational structure of single-plate graphical models, enabling implementation with CUDAHM.
In Sec.~\ref{sec:lum_func_sim} we present tests of such an implementation, using simulated data.
Sec.~\ref{sec:discussion} provides a summary and describes plans for future work.
The supplementary material includes an overview of the CUDAHM application programming interface, a derivation of the likelihood function for thinned latent point process models, and a description of the parametric luminosity distribution used for the simulation study.

%===============================================================================
\section{CUDAHM motivation and design}
\label{sec:design}

%-------------------------------------------------------------------------------
\subsection{Motivating problem structures}
\label{sec:motiv}

Suppose we observe $N$ members of a large population, with the observed members indexed by $i=1$ to $N$.
Each object (member) has a property or properties $\psi_i$ (a vector in multiple-property cases); we are interested in estimating the collection of properties, $\{\psi_i\}$, or their distribution, but cannot measure every component of $\psi_i$ with high precision.
Instead, for each object, we have observed data, $D_i$, that provide information about $\psi_i$.
% In the following, we sometimes use bold symbols to refer to quantities collectively, e.g., $\psivec \equiv \{\psi_i\}$, and $\Dvec \equiv \{D_i\}$.

We consider problems where the nature of the observations motivates models that specify a joint sampling distribution for the data, conditional on the member properties, that factors into a product of conditionally independent \emph{member sampling distributions},
\begin{equation}
p(\{D_i\} \assm \{\psi_i\}) = \prod_{i=1}^N p(D_i \assm \psi_i).
\label{data-joint-member}
\end{equation}
(For the sake of simplicity, we denote random variables and their values with the same symbol.)
We model the member properties, $\{\psi_i\}$, as IID draws from a \emph{population probability density function} (PPDF), $f(\psi_i;\theta)$, with uncertain parameters, $\theta$.
Goals of inference may include estimation of the PPDF (i.e., estimation of $\theta$), or estimation of the member properties.

\begin{figure}
\begin{center}
\includegraphics[width=.8\textwidth]{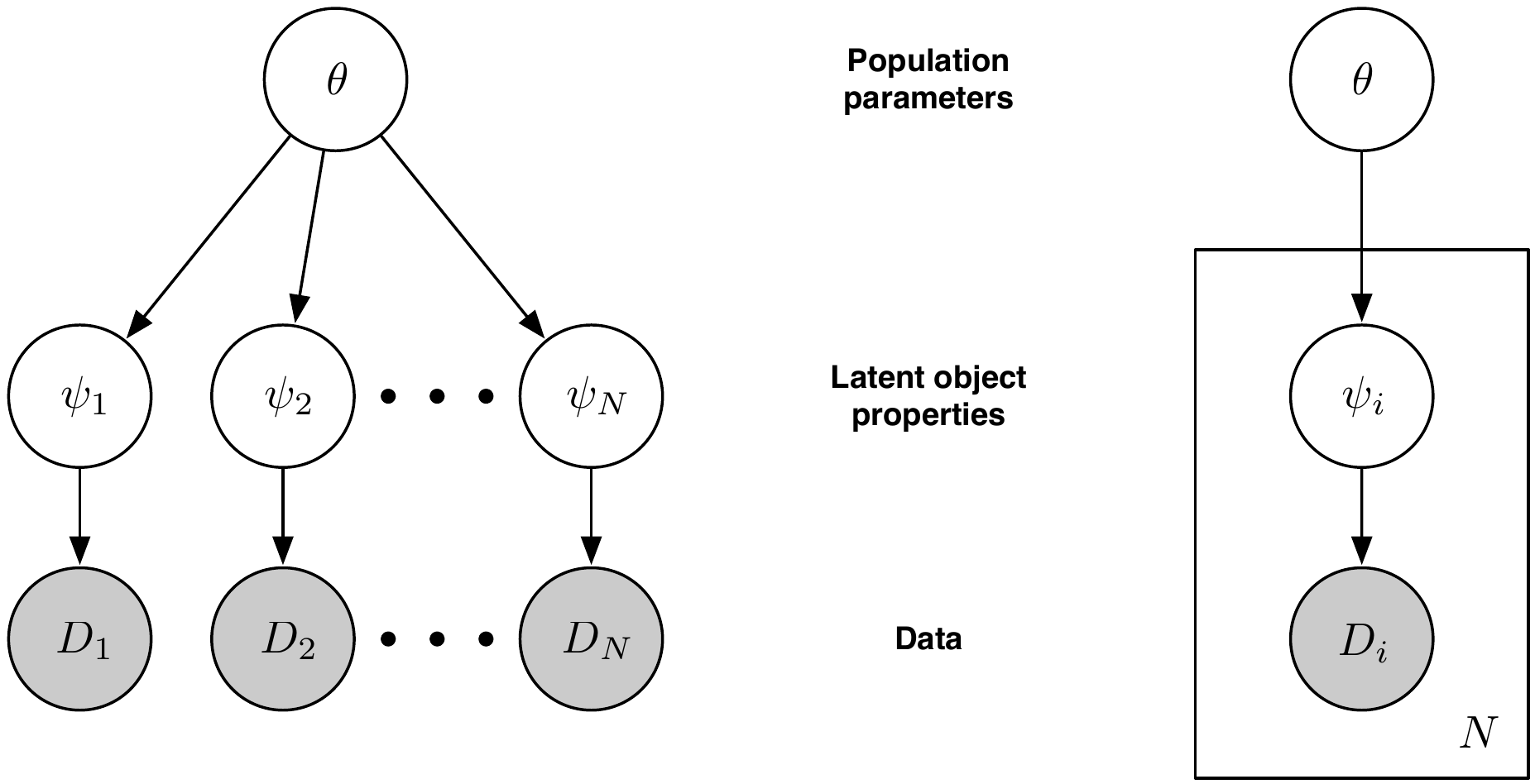}
\end{center}
\caption{Directed acyclic graph (DAG) for a 2-level hierarchical Bayesian model.
\emph{Left}:~DAG explicitly showing replicated conditionally independent subgraphs.
\emph{Right}:~DAG depicting replicated elements with a plate.}
\label{fig:DAG-2Level}
\end{figure}

Fig.~\ref{fig:DAG-2Level} shows the familiar DAG for this type of model, both explicitly and using plate notation.
Following common conventions, open nodes indicate uncertain random variables that are targets of inference, and shaded nodes indicate observed quantities, i.e., random variables that are uncertain a priori, but that become known after observation, and thus may be conditioned on.
If we denote the prior PDF for the population distribution parameters by $\pi(\theta)$, this DAG indicates that the joint PDF for all random quantities in this model may be written,
\begin{align}
p(\theta, \{\psi_i\}, \{D_i\})
  &= \pi(\theta) \prod_{i=1}^N f(\psi_i;\theta)\, p(D_i \assm \psi_i)\\
  &\propto \pi(\theta) \prod_{i=1}^N f(\psi_i;\theta)\, \mlike_i(\psi_i),
\label{joint-2level}
\end{align}
where we have defined the \emph{member likelihood functions},
\[
\mlike_i(\psi_i) \propto p(D_i|\psi_i).
\label{mlike-def}
\]
Note that, as likelihood functions (vs.\ sampling distributions), these functions need only be specified up to proportionality.
In particular, any dependence on $D_i$ that does not influence the dependence on $\psi_i$ can be ignored.

The DAG describes the conditional structure of a generative model for all of the random variables, including the data.
The model is completed by specifying the distributions for each node.
The bottom (data) nodes corresponds to sampling distributions for the data.
Traditionally---particularly, in probabilistic languages such as BUGS, JAGS, or Stan---the sampling distribution is explicitly specified.
However, when the task is inference of parameters conditional on observed data (versus prediction of unobserved data), specifying member likelihood functions (rather than sampling distributions) can be significantly more straightforward than specifying sampling distributions.
In many astronomical applications, estimation of $\psi_i$ from $D_i$ is often a nontrivial inference problem in itself.
For example, when $\psi_i$ denotes the apparent brightness of a star and $D_i$ denotes image data, inference may involve fitting the complicated point spread function of an imaging instrument to Poisson distributed photon counts in dozens or hundreds of pixels (often marginalizing over an uncertain background or instrument calibration component).
The resulting likelihood function for $\psi_i$ (or marginal likelihood function, when there are nuisance parameters) will often be relatively easy to summarize as a function of $\psi_i$; e.g., it may be well approximated by a Gaussian or multivariate Gaussian function (perhaps after a transformation).
On the other hand, the sampling distribution for the image data may be quite complicated and high-dimensional.
In many circumstances, it may not even be well-defined.
Weather or spacecraft conditions may affect the precision, accuracy, and even the quantity of data for a member observation; the repeated sampling distribution may be hard or even impossible to define objectively, while the likelihood function may be well-defined.
Most astronomical surveys produce member estimates with heteroscedastic uncertainties, in the sense of producing member likelihood functions with widths that vary from object to object.
It may be difficult or impossible to accurately describe the repeated sampling properties of the heteroscedastic uncertainties.
But for inference based on \emph{given} observations, only the actually available member likelihood functions matter.
Implementing inference in a manner that requires specifying only the member likelihood functions, rather than the sampling distributions, is a better fit to the nature of astronomical survey catalog data summaries than an implementation requiring unique specification of the lowest level sampling distributions.

A widely-used approach for posterior sampling in the context of two-level hierarchical models is the \textit{Metropolis-within-Gibbs} (MWG) algorithm, where the $\theta$ population parameters and the $\psi_i$ member properties are sampled in separate, alternating steps.
First, the member properties are sampled by holding the population parameters fixed, then the population parameters are sampled by holding the member properties fixed.
These steps may each be implemented with Metropolis or Metropolis-Hastings algorithms; their sequential combination amounts to Gibbs sampling on the joint space.
Explicitly, the steps are:
\begin{align}
\psi_i &\sim p(\psi_i \assm \theta, D_i), \quad\forall i\in 1:N;
\label{eq:psi_sampling} \\
\theta &\sim p(\theta \assm \psivec, M).
\label{eq:theta_sampling}
\end{align}
The departure point for CUDAHM is recognition that, since the $\psi_i$ properties are conditionally independent in \eqref{eq:psi_sampling}, they may be sampled in parallel, making this part of the MWG algorithm suitable for a massively parallel implementation using GPUs.
We describe such an implementation below.

\begin{figure}
\begin{center}
\includegraphics[width=.9\textwidth]{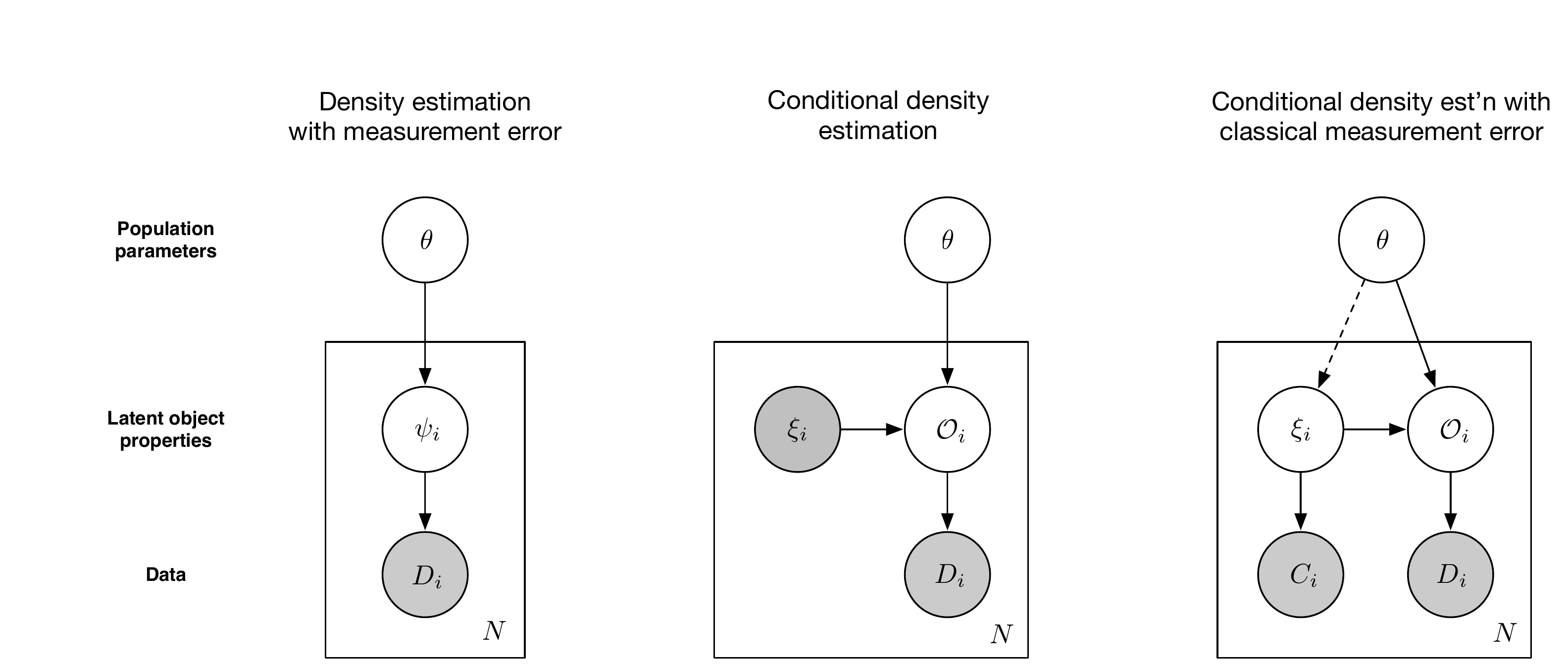}
\end{center}
\caption{Example single-plate DAGs that may be implemented in CUDAHM.
\emph{Left}: DAG for a 3-level hierarchical model corresponding to demographic inference for objects with latent characteristics $\chi_i$, related to latent observables $\obsv_i$.
\emph{Center}:~DAG for conditional density estimation, expressed via a latent observable $\obsv_i$, and a precisely measured predictor (covariate), $\xi_i$.
\emph{Right}:~DAG for conditional density estimation with classical measurement error, with a latent predictor, $\xi_i$, measured indirectly via data $C_i$.
The predictor may have an a priori known prior distribution, or it may be parameterized (with parameters included in $\theta$, in which case the dashed edge would be present).}
\label{fig:DAGs}
\end{figure}

Since $\psi_i$ may be a vector, the $\psi_i$ node in the DAG may admit a factorization leading to further structure within the plate in Fig.~\ref{fig:DAG-2Level}.
Fig.~\ref{fig:DAGs} shows DAGs for several other single-plate modeling scenarios for which inference may be implemented using MWG with massively parallel sampling of member properties.

The DAG in the left panel depicts a frequently arising structure in astronomy, where the object properties $\psi_i$ consist of intrinsic \emph{characteristics} $\chi_i$ that, if known, could predict \emph{observables}, $\obsv_i$, i.e., quantities that directly specify probability distributions for observed data.
(This terminology reflects astronomical parlance, where measurements of observables are used to ``characterize'' an object.)
An important example is inference of \emph{number-size distributions} (also known as number counts or $\log N$--$\log S$ distributions).
Here the object characteristics are distance, $r_i$, and luminosity, $L_i$ (amount of energy emitted per unit time).
The observable, $\obsv_i$, is flux (rate of energy flow per unit area normal to the line of sight, per unit time, at the telescope), which we denote by $F_i$.
It is related to the characteristics via the inverse-square law, $F_i = L_i/(4\pi r_i^2)$ (or its cosmological generalization).

The DAG in the middle panel depicts conditional density estimation, where the properties $\psi_i$ are comprised of precisely measurable predictors (covariates), $\xi_i$, that, together with the population parameters $\theta$, specify the PDF for observables, $\obsv_i$; the data provide likelihood functions for the $\obsv_i$.
An important example is inference of a \emph{luminosity function}, which describes the population distribution for the luminosities of a class of sources (say, a stellar or galaxy type).
If the PDF for luminosity is $\lpdf(L;\theta)$, and the distances to objects may be precisely measured (say, via spectroscopic redshift data), then by a simple change of variables the PDF for the flux observable for a source at distance $d$ is $4\pi d^2 \lpdf(4\pi d^2 F; \theta)$ (in Euclidean space).
This would be the distribution for the $\obsv_i$ node in the middle DAG.
We treat a more complicated version of this problem below, where the object sample is subject to flux-dependent selection effects.

As a final example, the DAG in the right panel depicts conditional density estimation with measurement error (i.e., with uncertainty in the predictors), with a classical measurement error structure (i.e., with the data distributions conditional on latent predictor values so the data provide ``noisy'' versions of the quantities in the density function).
A wide variety of astronomical data analysis problems have this structure.
\cite{K+12-DustSEDs} describes a noteworthy example studying how the spectrum of infrared emission from heated interstellar dust depends on properties of the dust grains.
This strongly nonlinear measurement error problem is one of the specific problems motivating CUDAHM.
Earlier studies, based on maximum likelihood estimates of dust properties (ignoring measurement error), found a surprising negative correlation between dust temperature and a spectral index parameter indicating how the dust properties tilt the infrared spectrum away from a black body spectrum.
Accounting for measurement error \emph{reversed the sign} of the inferred correlation, reconciling it with some theoretical models.
\cite{K+12-DustSEDs} analyzed measurements from $\sim\!10^4$ dust regions; CUDAHM dramatically accelerates the calculations and makes such studies feasible with 10 to 100 times larger samples.

%-------------------------------------------------------------------------------
\subsection{CUDAHM architecture}
\label{sec:arch}

To sample member propertis in the MWG algorithm, as specified in Eq.~\ref{eq:psi_sampling}, we use the robust adaptive Metropolis (RAM) algorithm devised by \cite{vihola2012robust}.
It works by adaptively refining a Metropolis algorithm proposal distribution during the sampling process until a target mean acceptance rate $\alpha_*$ is reached.
CUDAHM currently uses a multivariate normal distribution as the proposal $q$, and sets the target mean acceptance probability to a default value of $\alpha_{*}=0.4$.
Adaptation involves using new samples to adjust the proposal covariance matrix in a manner that decays with time along the Markov chain so as to guarantee correct asymptotic sampling.
Specifically, adjustments enter with a decaying weight, $\eta_{n}=n^{-2/3}$, where $n$ is the iteration number along the Markov chain.

Following \cite{vihola2012robust}, let $S_{1}$ be the identity matrix and $X_{1}$ some point in the space to be sampled for which the target density $\pi(X_{1})>0$.
Each RAM iteration cycles through the following steps: 
\begin{enumerate} \item Compute $Y_{n}=X_{n-1}+S_{n-1}U_{n}$, where $U_{n}\sim q$ is an independent random vector.
\item With probability $\alpha_{n} \equiv \min\{1,\pi(Y_{n})/\pi(X_{n-1})\}$ the step is accepted, and $X_{n}=Y_{n}$; otherwise the step is rejected and $X_{n}=X_{n-1}$.
\item Compute the lower-diagonal matrix $S_{n}$ with positive diagonal elements satisfying the equation
\begin{equation}
S_{n}S_{n}^{T}=S_{n-1}\left(I+\eta_{n}(\alpha_{n}-\alpha_{*})\frac{U_{n}U_{n}^{T}}{\parallel U_{n}\parallel^{2}}\right)S_{n-1}^{T}
\end{equation}
where $I$ is an identity matrix.
The solution for $S_{n}$ is unique as it is the left Cholesky factor of the right hand side.
\end{enumerate}
CUDAHM implements these steps on the GPU, via ``kernel'' code (code executed on a GPU) written in the CUDA language and taking advantage of optimized CUDA library functions (e.g., for pseudo-random number generation and linear algebra).
For the population sampling step in the MWG algorithm, as specified in Eq.~\ref{eq:theta_sampling}, parameter updates are computed on the host CPU, using code written in standard \Cpp, but with the log-posterior calculation (using the updated population parameters) executed on the GPU.

CUDAHM enables \Cpp\ programmers to rapidly construct an MWG sampler for a simple hierarchical model, requiring the user to supply only a minimimal amount of CUDA code.
For basic applications, the user must define two functions, instantiate a single class that manages the computations, and create a configuration file read by the compiled application that specifies parameters defining an MCMC run.
The functions compute the probabilities needed by the two steps in the MWG algorithm.
Appendix~A provides a high-level overview of the CUDAHM application programming interface, outlines the procedure for building a CUDAHM application, and provides details on accessing the source code repository, which includes code used for the computations reported below.

%===============================================================================
\section{Normal-normal performance test}
\label{sec:perform}

As a simple benchmark problem for comparing GPU and standard CPU computations for the same hierarchical model, we implemented GPU and CPU posterior samplers for the familiar normal-normal hierarchical model.
We consider a population of $N$ objects with three characteristics, denoted by vector latent member parameters $\chivec_i$ ($i=1$ to $N$) with components $\chi_{ij}$ ($j=1$ to 3).
The population distribution is multivariate normal with mean $\mupop$ and covariance matric $\covpop$, and the measurements $\yvec_i$ (with components $y_{ij}$) have uncorrelated, heteroscedastic zero-mean measurement errors with variances $\sigma_j^2$.  
That is,
\begin{gather}
\chivec \sim \MVN(\mupop, \covpop),\\
y_{ij} \sim \Norm(\chi_{ij}, \sigma_j^2).
\end{gather}
We simulated data using 
\begin{equation}
\mupop = 
  \begin{bmatrix*}[r]
  1.2 \\
  -0.4 \\
  3.4
 \end{bmatrix*},
\qquad
\covpop = D\cdot R\cdot D,
\end{equation}
where $D = \diag(2.3, 0.45, 13.4)$ is a diagonal matrix of (marginal) population standard deviations for the components of $\chivec$, and $R$ is their correlation matrix, with
\begin{equation}
\begin{bmatrix}
  1 & 0.3 & -0.5 \\
  0.3 & 1 & 0.54 \\
  -0.5 & 0.54 & 1
 \end{bmatrix}.
\end{equation}
For the measurement error scales, we used $\sigma_j = (1.2, 0.4, 0.24)$.
This represents a population with strongly correlated characteristics, with population standard deviations varying by a factor of $\approx 30$.
Thus a posterior sampler would have to adapt to very anisotropic distributions for the many latent parameters.
Also, the measurement error is comparable to the population dispersion for two of the characteristics, so the calculation is implicitly doing a significant amount of deconvolution.

For inference, we considered $\covpop$ and $\sigma_j$ to be known, and sought to estimate the population means, $\mupop$, using uniform hyperpriors.

% Not sure where this came from:
% We set the prior on $\mupop$ to be multivariate normal with zero mean and covariance equal to $\diag(50, 5, 1000)$, which is relatively noninformative for the first and third $\chivec$ components, but somewhat informative for the second.

% GTX 560 specs:
% https://www.geforce.com/hardware/desktop-gpus/geforce-gtx-560/specifications

We implemented this model using CUDAHM, running on a desktop computer with single NVIDIA GeForce GTX~560 GPU card, a relatively inexpensive consumer graphics card; this demonstrates capability that can be achieved with limited hardware expense.
This card has 336 CUDA cores and 1~GB of memory; each core can run many concurrent threads of computations (with a total thread count $\approx 5000$).
For comparison, we implemented the model in the Stan probabilistic programming language (via the PyStan Python interface), running the calculations on a quad-core 2.66~GHz Intel Xeon CPU with 16~GB of memory.
We used Stan's default algorithm, an automatically tuned Hamiltonian Monte Carlo (HMC) sampler, with four chains run in parallel across the four cores.
Of necessity, any GPU/CPU comparison will be an ``apples-to-oranges'' comparison.
Here, our thinking was to compare against a sophisticated, adaptively tuned algorithm that, despite its sophistication, is easy to use, thanks to the high-level nature of the Stan language.
In comparison, it is significantly more time consuming to implement this model in \Cpp\ using CUDAHM.
We wanted to see what performance gains a user gets for the extra work.
Note that, despite the simplicity of the Stan language, Stan creates a \Cpp\ library on-the-fly implementing the specified model, so in this comparison the CPU implementation is not compromised in performance by an interpreter or virtual machine layer.

We followed conventional practice of throwing out early samples, comprising a ``burn-in'' period, to avoid initialization bias, using only post-burn-in samples for inference.
Different algorithms and initialization policies affect the burn-in duration in complicated ways.
To compare performance, we wanted to focus on the rate of producing samples once a sampler was mixing well.
We chose to compare how long it took to produce a sample with an effective sample size (ESS) of $\approx 1000$, \emph{after} burn-in, using the smallest ESS for the $\mupop$ parameters.
We were motivated here again by astronomical applications, where scientific interest usually focuses on population properties, and not on improving estimates of latent member parameters (though we did verify that the $\chivec$ parameters were well-sampled).

To avoid influence of initialization bias on posterior inferences, Stan adopts a default burn-in policy of throwing out the first half of the requested number of samples, and computes the potential scale reduction statistic, $R$, from the parallel chains, for assessing convergence.
We ran chains of length 500, producing $4\times 250 = 1000$ final samples; $R$ was within a few percent of unity in all cases, indicating sufficient burn-in.
The reported ESS for the population parameters ranged from 500 to 1000 in our runs, indicating very rapid mixing of the chains after burn-in.
It should be noted that the HMC algorithm generates proposals by solving a differential equation, producing a proposed step by evolving the previous point deterministically along a path in parameter space (with random initial conditions).
Thus each sampling iteration requires multiple evaluations of the likelihood function and its gradient with respect to all parameters, so the number of evaluations of the likelihood function is much larger than the combined number of steps in the chains.
In addition, tuning of the algorithm (during the burn-in period), while automatic, can be computationally much more expensive than sampling with the optimized algorithm, so that if a modest number of posterior samples is sought, the majority of time is spent tuning rather than sampling.

For CUDAHM, we did a trial run (for each population size) and examined trace plots and autocorrelation plots to graphically estimate a conservative burn-in time.
For the range of population sizes explored here ($10^2$ to $10^4$), the burn-in length ranged from $4\times 10^4$ to $10^5$.
We ran chains for $7.5\times 10^4$ steps after burn-in.
The RAM-based MWG sampler produces strongly correlated samples, and the typical ESS was $\approx 1100$ to 1500.

\begin{figure}[t]
\begin{center}
\includegraphics[width=.9\textwidth]{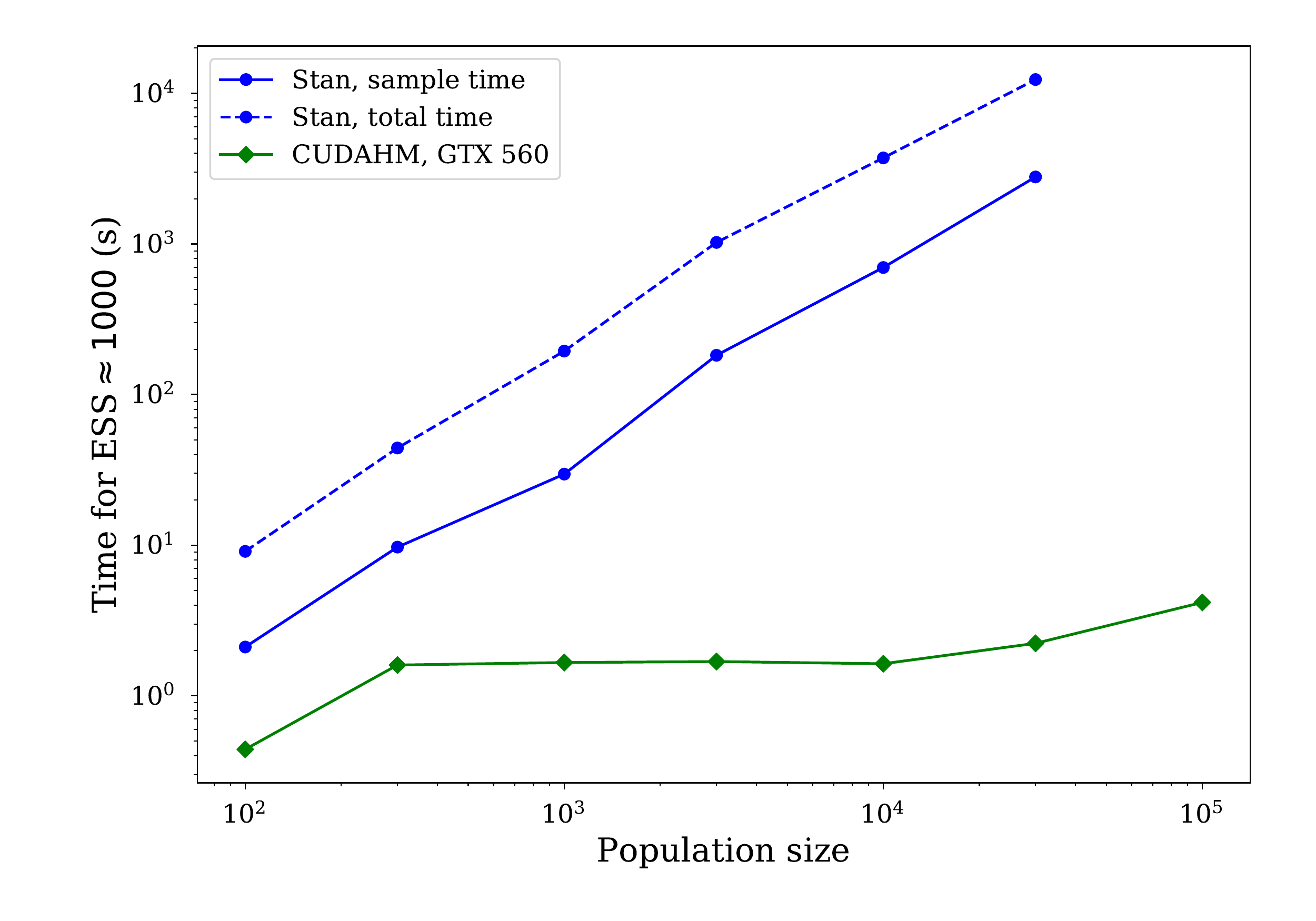}
\end{center}
\caption{Performance of CUDAHM and Stan implementations of a simple normal-normal benchmark hierarchical model, measured by the effectively independent sample generation rate, shown as the time needed to achieve ESS of $10^3$ for the three population parameters.
For Stan, the dashed curve additionally shows the total time (burn-in plus sampling) to get ESS$=1000$.}
\label{fig:NNPerf}
\end{figure}

We recorded the post-burn-in run times and ESS values, allowing us to estimate ESS rates (for Stan we used the run time of the fastest chain, presuming slower chains were more affected by background processes).
Fig.~\ref{fig:NNPerf} shows the rates as the time needed to produce a chain with effective sample size (ESS) of 1000, as a function of population size.
For Stan, there is a clear scaling behavior, with time proportional to $N^{\alpha}$ with $\alpha\approx 1.27$.
The computational complexity of MCMC algorithms is complicated even for simple algorithms (\citealt{BA09-CompComplexMH}); we are not aware of theoretical work explaining the observed behavior.
The figure also shows the \emph{total} run time (burn-in plus sampling) for the Stan runs.
For the modest-length runs used here, tuning dominates the computational expense by a significant factor $\sim 4$.
This is a fixed cost, so if a large ESS is sought, the burn-in time may become negligible.

%The approximately constant cost is presumably due to some kind of fixed-cost overhead involving communication between the CPU and GPU (likely exacerbated because the GPU card is sharing display duties).

For CUDAHM, the rate appears nearly constant up to a population size of $10^4$.
The delayed onset of increasing cost with respect to population size is likely due to the size of the GPU thread pool, with cost not increasing until the thread pool is filled.
For population sizes greater than $10^4$, we can begin to see the computational cost of posterior exploration dominating fixed costs.
(We were unable to run calculations with larger populations on this hardware due to limited memory on the CPU side.)
For the small populations, CUDAHM is over an order of magnitude faster than Stan.
By a population size of 500, CUDAHM is 10 times faster than Stan at exploring after burn-in; for $N=3\times10^4$, it is over 1000 times faster.
(For desired posterior sample sizes of order a few thousand, Stan's long burn-in/tuning time makes the factor two to three times larger.)

This example demonstrates that massive parallelization of an algorithm of modest complexity can achieve much greater performance for hierarchical inference than a few-core parallel implementation of a complex algorithm.
Recently in this journal, \cite{B+16-HMC+GPU} demonstrated very significant GPU acceleration of HMC for multinomial regression.
They used a high-level GPU interface available in Python, which has restricted application in comparison to writing CUDA code directly in C or \Cpp.
They speculated that HMC for standard hierarchical models could be implemented within the restrictions of the Python GPU interface.
Our benchmark suggests this could lead to enormous speedups for such models, perhaps even greater than we find with our RAM-based MWG algorithm.
CUDAHM was implemented in \Cpp\ to enable application to models outside of standard setups, including nonlinear models with complex measurement error structure.
We turn to such an example in the next section.

%===============================================================================
%\section{Example: Dust properties in a star-forming region}
%\label{sec:dust}
%
%\enote{Add material here for Brandon's dust SED example if desired; we should probably try to save this for the A\&C version.}
%===============================================================================

% These perhaps should be merged:
%===============================================================================
\section{Luminosity function estimation: Thinned latent point process framework}
\label{sec:lum_func}

As a moderately complicated application of CUDAHM, we here consider luminosity function estimation, a parametric conditional density estimation problem arising across many areas of astronomy.
We highlight this example, because of its ubiquity across astronomy, because it has some statistically unique features, and to illustrate the generality of CUDAHM.
We describe luminosity function estimation for a model including not only measurement error, but also \emph{selection effects}.
The selection effects make the joint distribution structure more complicated than the conditional density estimation DAG shown in Fig.~\ref{fig:DAGs}; in particular, there are two plates (corresponding to detected and undetected objects).
However, the likelihood function can be manipulated to have a conditional independence structure similar to that of a single-plate DAG, allowing straightforward implementation of the model using CUDAHM.

%-------------------------------------------------------------------------------
\subsection{Astronomical background}
\label{sec:LF-astro}

The fundamental observables for localized astronomical sources include position (both direction on the celestial sphere, and distance, $r$, in some chosen coordinate system), and apparent brightness, quantified in terms of flux, $F$ (astronomers also often report brightness using magnitudes, a negative logarithmic relative flux scale).
Astronomers use these observables to study demographic properties of many classes of sources, including stars and galaxies of various types, minor planets (such as asteroids), and explosive transients (gamma-ray bursts, supernovae).
For concreteness, here we focus on observations of nearby galaxies, for which distance may be measured by using spectroscopy to find the \emph{redshift}, $z$, of spectral lines (i.e., their fractional shift in wavelength from laboratory values).
Due to the cosmological expansion, for relatively nearby galaxies outside the local group, distance is proportional to redshift to a good approximation, with
$r = cz/H_0$,
where $c$ denotes the speed of light, and $H_0$ is Hubble's constant, measuring the current expansion rate of the universe, with $H_0 \approx 70$~km~s$^{-1}$~Mpc$^{-1}$ (with Mpc denoting megaparsecs).
$H_0$ is measured with a precision of several percent, and spectroscopic redshifts for nearby galaxies can be measured to sub-percent precision.
For simplicity, here we consider distances to be precisely measured, via spectroscopic redshifts.
Often, astronomers use redshift directly as a proxy for distance.

% Q: Mention photo-z as similar but with measurement error.

As noted previously, a fundamental intrinsic characteristic of a source is its luminosity, $L$, a measure of its intrinsic (vs.\ apparent) brightness, related to flux and distance via the inverse-square law.
The \emph{luminosity function}, $\lfunc(L, r)$, describes the distribution of luminosities for a population at a specified distance (or redshift).
It is typically defined as the intensity function for a point process, i.e., as specifying the expected number of galaxies per unit volume at distance $r$, per unit luminosity interval.
Equivalently, we can consider the luminosity function to define a marked point process, with an intensity function for spatial location, and a PDF for luminosity (considered as a mark).
If we denote the spatial number density of galaxies at distance $r$ by $n(r)$ (galaxies per unit volume), then $n(r) = \int dL\, \lfunc(L, r)$.
The \emph{luminosity PDF} for galaxies at distance $r$ is then
\begin{equation}
\lpdf(L,r) = \frac{\lfunc(L, r)}{n(r)}.
\label{eq:lpdf}
\end{equation}
Note that $\lfunc(L,r)$ and $\lpdf(L,r)$ specify \emph{conditional} distributions, i.e., distributions for $L$ at a given $r$.
(Authors vary on the definition of the luminosity function, many defining it as done here, and others using ``luminosity function'' to denote what we here call the luminosity PDF.)
As noted above, the \emph{flux PDF} for galaxies at distance $r$, denoted $\rho(F,r)$, can be found by a change of variables from $L$ to $F$, giving
\begin{equation}\label{eq:fluxPDF}
\rho(F,r) = 4\pi r^2 f(4\pi r^2 F, r).
\end{equation}
Note that $\rho(F,r)$ is a conditional PDF for $F$, conditioned on $r=cz/H_0$.

% \noindent\enote{Define the distance PDF, $h(r)$, here?  Or leave it below?}

The galaxy luminosity function carries valuable information about the formation and evolution of galaxies, therefore it is an important target of inquiry in astronomy (see \citealt{BST88-LumFuncReview} and \citealt{J11-LumFuncReview} for reviews).
\cite{J11-LumFuncReview} provides a review of methods developed by astronomers for estimation of galaxy luminosity functions.
A nontrivial challenge is careful accounting of both measurement error and selection effects, which are related for most survey datasets.
This section outlines how these complications can be modeled (a detailed derivation of the likelihood function used here is provided in the Appendix of \cite{LH19-MultilevelHBCosPop-TLaMPP}).
Fundamentally, the model does not have the simple single-plate DAG structure that motivated the design of CUDAHM.
However, the joint posterior may be manipulated into a form well-suited to CUDAHM.
Luminosity function inference thus provides an example of the flexibility of the framework.
The next section presents a simulation study demonstrating CUDAHM luminosity function inference for simulated populations with sizes up to $10^6$.

% For simplicity we here focus on a homogeneous population, so that $n(r)$ is considered fixed (constant with respect to volume measure).
% The following development is straightforward to generalize to account for models with distance dependence in the luminosity function.

%-------------------------------------------------------------------------------
\subsection{Modeling survey selection effects and measurement error}
\label{sec:slxn+err}

Astronomers estimate luminosity functions and other astronomical distributions using data compiled in \emph{survey catalogs}: tables of estimates (including uncertainties) of object properties, accompanied by a description of selection criteria for the survey that produced the catalog.
Catalogs are derived data products; they summarize information in more complex and voluminous raw datasets, such as large collections of images or time series.
The nature of astronomical catalog data makes their analysis differ in some respects from analyses of survey data familiar in the statistical survey sampling literature.

Flux measurements are affected by measurement error that is often dominated by Poisson fluctuations in the photon counting rate, including fluctuations from astrophysical and instrumental backgrounds.
The measurement error thus approximately scales with the square root of the total flux (source plus background), and is fractionally greater at low flux than at high flux.
At low fluxes, photons from the backgrounds can produce false source detections.
To prevent this, surveys adopt detection criteria to strongly mitigate against false detections.
A simple, representative criterion is to accept sources only if the estimated flux is $\nu$ times the flux uncertainty, with $\nu \approx 5$ so that the probability for false detection is low even for large catalogs (i.e., there is strong control of the family-wise error rate).
Here we assume we are in the pure catalog regime, i.e., that the threshold is set high enough that the resulting catalog may be considered to have no false detections.
(The framework described below is straightforward to generalize to settings where false detections may be present, but most catalogs adopt high thresholds so we focus on the simpler pure catalog regime here.)

Detection criteria introduce \emph{selection effects} into catalogs.
Most obviously, faint sources (low luminosity sources, or distant high luminosity sources) are excluded; the observable luminosity function is a thinned version of the actual luminosity function.
In addition, measurement error more subtly but significantly distorts the shape of the observable distribution, a phenomenon well known in the density deconvolution literature, and also recognized in the astronomical literature, where it is sometimes called \emph{Eddington bias}, in reference to early discussions of the distortion by Eddington and Jeffreys (\citealt{J38-EddBias,E40-EddBias}).
They noted that an object with a measured flux of $\hat F$ is more likely to be an object with a true flux $F < \hat F$ than one with $F > \hat F$, even when measurement errors have a symmetric distribution, because dim sources are more numerous than bright sources in most astronomical settings.
Selection effects can exacerbate the distortion in the vicinity of a flux threshold, with measurement error and the falling flux distribution conspiring to scatter more below-threshold sources into the observed sample than above-threshold sources out of it, a component of what astronomers call \emph{Malmquist bias} (\citealt{LH10-BMIC}; but note that the term is used inconsistently in the literature).
Hierarchical modeling can automatically account for such thinning and distortion, in a manner that adapts to the shape of the luminosity function; this is a major motivation for its increasing popularity in astrostatistics.

We have developed a framework for modeling astronomical survey data using \emph{thinned latent marked point process models}.
Measurement error is handled in a hierarchical Bayesian manner, by introducing latent member property parameters, with catalog estimates understood as describing member likelihood functions for the properties.
We model the population distribution of the latent member properties using marked point processes.
We model selection effects through thinning of the latent point process.
This framework was originally developed for studying the luminosity function of gamma-ray bursts, powerful explosive transients thought to mark the birth of stellar-mass black holes (\citealt{LW95-GRBs-TLPP,LW98-GRBs-Iso}).
It was subsequently adapted to study the luminosity distribution (and through it, the size distribution) of trans-Neptunian objects, asteroid-like minor planets in the outer solar system (\citealt{L04-MsmtErr,P+08-TNOSizeDistn}).
We outline the framework here as it applies to luminosity function estimation and similar problems, in contexts where a marked Poisson point process is an appropriate model for the latent member properties.
\cite{kelly2008flexible} independently developed a similar approach for settings where a binomial point process may be appropriate, and applied it to estimating the number density of quasars as a function of redshift.

A somewhat subtle aspect of the problem is the tie between measurement error and selection effects.
In astronomical surveys, the raw data are searched to find candidate objects.
For candidates that pass detection criteria, the data are used to estimate source properties.
The same underlying data are used both for selection (detection) and measurement; these tasks are not independent, as is assumed in many statistical survey methods outside of astronomy.
Ignoring the dependence can corrupt inferences.

%Modeling selection effects requires considering the properties of objects that were excluded from the catalog, i.e., objects that were present but that produced data that did not meet detection criteria.

Loredo \& Hendry (\cite{LH19-MultilevelHBCosPop-TLaMPP}, see the Appendix) provide a derivation of the likelihood function for the parameters of a luminosity function or luminosity PDF, based on catalog data from a survey subject to measurement error and selection effects in fluxes, and reporting precise distance measurements, $r_i$ (see \cite{L04-MsmtErr} and references therein for earlier, simplified treatments).
Denoting the luminosity PDF parameters by $\rhopar$, the likelihood function takes the form
\begin{equation}\label{eq:rho-mlike}
\like(\rhopar)
  = \prod_{i=1}^N \int \dd F\, \epdf(F,r_i;\rhopar)\, \ell_i(F),
\end{equation}
where we have introduced an \emph{effective density}, $\epdf$, for the latent observables, $F$ and $r$, related to the luminosity PDF, $\rho(F,r;\rhopar)$, according to
\begin{equation}\label{eq:epdf}
\epdf(F,r;\rhopar) \equiv
  \frac{h(r) \rho(F,r;\rhopar)}
    {\int \dd r \int \dd F\,\aeffic(F)\, h(r)\, \rho(F,r;\rhopar)},
\end{equation}
where $h(r)$ is the distance distribution (assumed uniform in volume, and thus $\propto r^2$ as a function of distance), and $\aeffic(F)$ is the average detection efficiency for sources with flux $F$ (the detection efficiency at each location in the images comprising the survey, averaged over location).
Equation~(\ref{eq:rho-mlike}) resembles the likelihood function for a binomial point process in flux and distance, but with observations that have measurement errors described by the member likelihood functions.
However, the analogy is not exact, because the effective density is not a PDF in $(F,r)$---the denominator in \eqref{eq:epdf} is not the integral of the numerator (because of the $\aeffic$ factor).
This is not an inconsistency, because the likelihood function is not a probability distribution for $(F,r)$, rather, it is a probability distribution for the data (up to proportionality).

Although this model does not exactly match the simpler model structures discussed previously, the structure of the likelihood function in (\ref{eq:rho-mlike}) is essentially the same as that for conditional density estimation with measurement error (the middle DAG in Fig.~\ref{fig:DAGs}).
It has a single product term, composed of independent factors for each detected object---just the type of structure CUDAHM was designed to sample from.

%===============================================================================
\section{Luminosity function estimation: Simulation study}
\label{sec:lum_func_sim}

To explore the performance of CUDAHM for luminosity function estimation, we implemented the framework just described and applied it to simulated data.
To focus on performance, this example ignores important complexities arising in modeling real galaxy catalog data.
For example, we ignore cosmological corrections to the inverse square law (which depend on uncertain cosmological parameters); these are important for samples containing distant galaxies.
We also ignore diversity in the spectra of galaxies.
This is important for real data because instruments gather light in limited spectral ranges, determined by properties of the atmosphere, telescope optics, and detector sensitivity.
Among the optical elements are filters that deliberately restrict the spectral passband, so that repeated measurements with different filters can provide broad-band information about galaxy spectra.
As a result, galaxies with the same bolometric (full-spectrum) luminosity and distance, but different spectral shapes, will have different apparent brightnesses (fluxes).
A full analysis would incorporate data from multiple bands.
For the study described here, we assume the simulated galaxies have the same spectra, or that the catalog estimates have been adjusted for spectral diversity.

%................................................................................
\subsection{Simulation setup: population distribution and priors}
\label{sec:simsetup-popn}

For our simulation study, we adopt an \emph{exponential-cutoff break-by-one} (BB1) luminosity function that has a smoothly broken power law behavior for luminosities below a characteristic scale, and an exponential cutoff for larger luminosities.
For astrophysically relevant parameters it has a monotonically declining PDF, i.e., low-luminosity power law dependence with a negative exponent.
It may be seen as a generalization of the gamma distribution, and of a  model popular in astronomy called the Schecter function (essentially a gamma distribution with a small shape parameter).
The motivation for the BB1 model, and some detailed properties, are presented in Appendix~C.
  
The BB1 model has a luminosity PDF with $\theta$ composed of three parameters: a mid-luminosity nominal power law index, $\beta$, and two parameters defining the mid-luminosity range, $(l, u)$, with $l < u$.
For $l<L<u$, the behavior is approximately $\propto L^\beta$.
As $L$ decreases below $l$ the power law behavior smoothly changes, with the local power law index increasing to $\beta+1$ at low luminosities.
At high luminosities (above $u$), the PDF falls exponentially  ($u$ thus plays the role of the $L_*$ parameter known to astronomers in the Schecter function).
The BB1 luminosity PDF has following functional form:
\begin{equation}
\label{eq:lumPDF} 
\lpdf(L ; \theta) = 
  \frac{C(\beta,u,l)}{u}\left(1-e^{-L/l}\right) \left(\frac{L}{u}\right)^{\beta} e^{-L/u},
\end{equation}
where the normalization constant $C(\beta,u,l)$ is given in Appendix~C.
Note that as $l\rightarrow 0$, the BB1 distribution becomes a gamma distribution (if $\beta > -1$).
%Also, in our computational implementation, the condition $\beta=-1$ of the first case is $-1.001<\beta<-0.999$.
We designed the BB1 distribution to have smooth power law break behavior at low $L$, yet also have an analytical normalization constant;
it is proper for $\beta > -2$.
It can also be sampled from using a straightforward modification of a widely-used algorithm for sampling from the gamma distribution (\citealt{ahrens_computer_1974}).
These properties make it useful for simulation experiments.

% We define a BB1 luminosity function by multiplying the BB1 luminosity distribution by the galaxy spatial number density, which is simply a constant, $n$, for a homogeneous population.

\begin{figure}
	\begin{subfigure}[c]{0.45\textwidth}
		\includegraphics{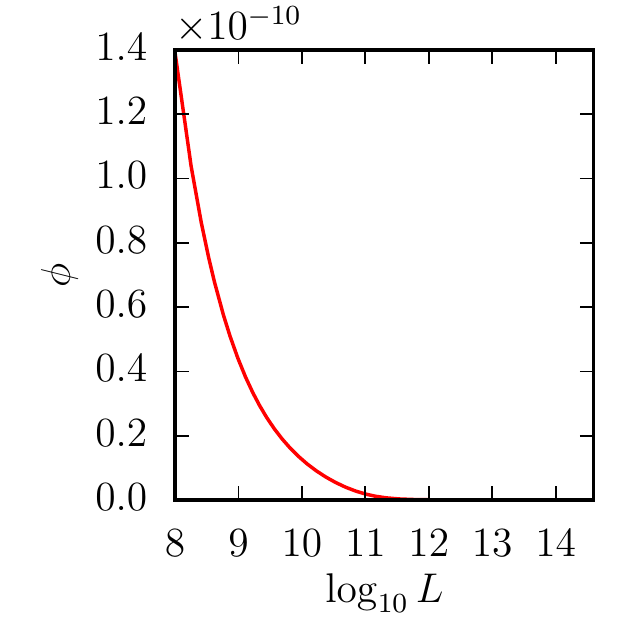}
		\caption{}
	\end{subfigure}
	\begin{subfigure}[c]{0.45\textwidth}
		\includegraphics{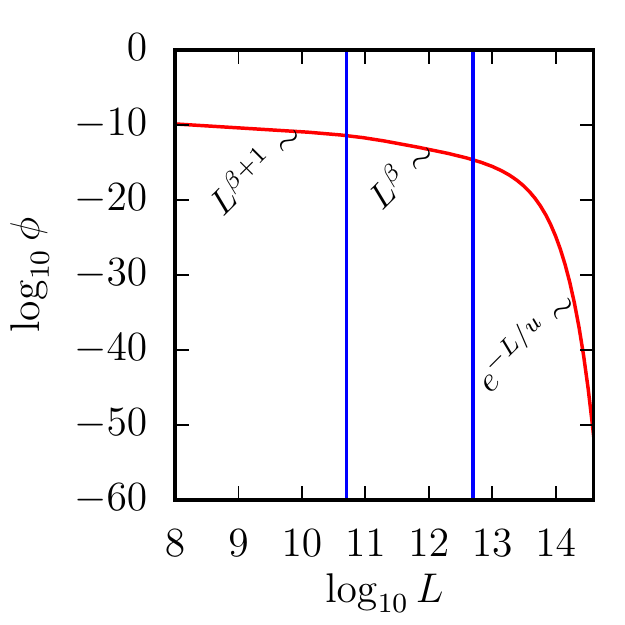}
		\caption{}
	\end{subfigure}
	\caption{Example BB1 truncated broken power law PDF, on a log-linear scale (a) and on log-log scale (b), with parameters $(l,u) = (5\times 10^{10}, 5\times 10^{12})$ in dimensionless units, and $\beta = -1.5$.
	In Panel~b the blue verticals show the lower limit $l$ and upper limit $u$ of the region where the power law slope is $\approx\beta$.}
	\label{fig:lumfunc}
\end{figure}

Fig.~\ref{fig:lumfunc} shows an example BB1 luminosity PDF, with $\beta = 1.5$, $(l,u) = (1\times 10^{8}, 1\times 10^{10})$ in solar luminosity ($L_\odot$) units; it is plotted both with $\log$-linear axes, and with $\log$-$\log$ axes, where the varying power law behavior is evident.

% TODO: Switch fig to new params

We simulate observations of a population described by the BB1 model, with parameters chosen so that galaxies with $L>l$ have a distribution similar to that found in the analysis by Blanton et al. (\citealt{blanton2003galaxy}, B03) of $\approx 150,000$ galaxies with spectroscopic redshifts observed in the Sloan Digital Sky Survey (SDSS).
We set $u = 1\times 10^{10} L_\odot$ (where $L_\odot$ denotes the solar luminosity; this $u$ value is approximately equal to the B03 value of $L_*$ in a Schecter function fit), $\beta = -1.5$ (slightly steeper than the B03 value), and $l= 1\times 10^{8} L_\odot$, corresponding to the lowest luminosities studied by B03.
We choose survey parameters corresponding to a deeper survey than SDSS, thus probing luminosities dimmer than $L=l$.
These choices are motivated in part by current and emerging surveys, such as the photometric (broad-band) surveys by the Panoramic Survey Telescope and Rapid Response System (Pan-STARRS, current) and the Large Synoptic Survey Telescope (LSST; starting in 2023),   and the spectroscopic (redshift) survey by the Dark Energy Spectroscopic Instrument (DESI; starting in 2019).

For distances, we sample values from a spatially homogeneous population extending out to a maximum distance $\rmax = 1$~Gpc.
Only very luminous sources can be detected from large distances.
For the detection criteria we adopt (see below), this maximum distance is such that galaxies are visible beyond $\rmax$ only if $L \gtrsim 20u$, an event with negligible probability (which we formally exclude by truncation).

% suppression factor  $e^{-20} \sim 2\times 10^{-9}$

% This appears to be an implementation detail: don't bother sampling L's you know you'll reject
% \enote{Janos also truncated below $l$; is this necessary?}

The upper luminosity scale, $u$, and maximum distance, $\rmax$, together define a fiducial flux value,
\begin{equation}\label{eq:Ffid}
\Ffid
  \equiv \frac{u}{4\pi r_{\rmax}^2} 
  \approx 3.2\times 10^{-13} \left(\frac{u}{10^{10} \Lsol}\right)
      \left(\frac{\rmax}{1~\text{Gpc}}\right)^{-2}\;
      \text{erg cm$^{-2}$ s$^{-1}$}.
\end{equation}
This is a convenient unit in which to express fluxes.
Although $\Ffid$ is minuscule, modern survey telescopes would detect $\sim 10^4$ to $10^5$ photons from a source with this flux.

% TODO:  Double-check this:
%\enote{Note the parameter value changes!  
%The plots need revision; luminosity axis labels need to be shifted by $\times 100$, and the axis label should read $\log_{10}(L/L_\odot)$.  
%Check the $\rmax$ value and make sure it corresponds to the flux limit and luminosity truncation after shifting to the new params.}

Now we consider the choice of prior for the population parameters.
For $\beta$, we adopt a prior that is flat with respect to the angle in $\log$-$\log$ space.
This choice has the virtue of not putting a lot of prior probability on the steep slope range, which a flat prior on $\beta$ would do.
Denoting the angle by $\varphi$, if we adopt a prior PDF of $g(\varphi)$ on $\varphi$, the prior on $\beta = \tan\varphi$ is $p(\beta) = g(\varphi)/(1 + \beta^2)$.
For a flat $\varphi$ prior between two cut-offs $\varphi_L$ and $\varphi_U$, this gives
\begin{equation}
	p(\beta) = \frac{1}{\varphi_U - \varphi_L} \cdot \frac{1}{1 + \beta^2}.
\end{equation}
This is a truncated Cauchy distribution.
The BB1 distribution requires $\beta > -2$, corresponding to $\varphi_L = -1.107$.
If we require Eq.~\ref{eq:lumPDF} to be decreasing, the upper limit becomes $\beta < 0$, corresponding to $\varphi_U = 0$.
Using these limits, the prior on $\beta$ is
\begin{equation}
p(\beta) = \frac{0.903}{1 + \beta^2} \quad \quad \quad \textrm{for} -2 < \beta < 0.
\end{equation}
For the upper scale we use a log-flat prior, a conventional choice for a scale parameter that must be positive.
Even though this will be improper on both sides, we can ignore the impropriety and the normalizing constant since the likelihood function will make the posterior proper.
A prior flat in $\log{u}$ corresponds to $p(u) \propto \frac{1}{u}$.
The lower scale $l$ must be below the upper scale $u$, which we can ensure by using a prior factored as $p(l, u) = p(u) p(l \assm u)$ and taking $p(l \assm u)$ to vanish for $l \geq u$.
A log-flat prior also seems appealing for $l$ but the data (i.e. luminosity measurements) do not probe the distribution down to zero due to the flux limit of the telescope, we could not have a proper prior without introducing a lower cut-off.
Instead, we simply use a flat prior on $l$.
Hence, the overall prior will be \begin{equation}\label{eq:popParsPriorPDF} p(\beta, l, u) \propto \frac{l}{u \cdot (1 + \beta^2)} \quad\quad\quad \textrm{for} -2 < \beta < 0, l < u.
\end{equation}

%................................................................................
\subsection{Simulation setup: detection and measurement}
\label{sec:simsetup-data}

We simulate measurement errors and selection effects using a simplified model commonly adopted in astronomical simulation studies (\citealt{F99-SDSSSim,LSSTRefDesign}).
Modern astronomical optical detectors, such as cameras using charge-coupled devices (CCDs), count photons.
The Poisson distribution accurately describes photon arrival and detection, but there are additional contributions to measurement uncertainty, including from backgrounds and electronic noise.
To screen out false detections, only reasonably strong candidate sources are accepted as genuine, so that catalog data are typically in the large-counts regime.
Most simulation models work in this regime, approximating the Poisson distribution by a normal distribution, and treating the additional contributions to measurement uncertainty also using normal distributions.
The overall measurement uncertainty thus is approximately normal, with a variance found by adding the variances of the component processes.

For simulation studies of hierarchical Bayes approaches, it is important to distinguish approximations of the sampling distribution used to generate simulated data, and approximations of the member likelihood functions needed for inference with a particular simulated dataset.
As a concrete illustration of the distinction, consider a source with true flux $F$ being measured by an ideal photon counting detector, with measurement uncertainty due only to Poisson counting uncertainty associated with the source flux.
For an instrument with projected area $A$ observing for a time $T$, the expected number of photons is $\mu = ATF$, and the standard deviation in the number of photons is $\mu^{1/2}$.
In the high-counts regime, we could simulate an observation by drawing a number of photon counts, $n$, from a normal distribution $N(\mu,\mu)$, i.e., with
\begin{equation}\label{eq:PoissonNormal}
p(n|F) \approx g(n;F) \equiv \frac{1}{(ATF)^{1/2}\sqrt{2\pi}}
  \exp\left[-\frac{(n-ATF)^2}{2ATF}\right].
\end{equation}
Suppose the simulated value of $n$ is $n_{\rm obs}$.
For analysis of that observation, we would need to approximate the member likelihood function based on that datum.

One possibility is to take $\mlike(F) \propto g(n_{\rm obs};F)$, that is, to make the member likelihood proportional to the Gaussian approximation of the \emph{sampling} distribution for the observed counts, $n_{\rm obs}$.
Note that this function is \emph{not} a Gaussian with respect to $F$.

Alternatively, we may seek an approximate member likelihood that is a Gaussian function of $F$.
The exact likelihood function, $\mlike(F)$, based on the Poisson sampling distribution, is proportional to a gamma distribution with shape parameter $\alpha = n_{\rm obs} + 1$ and scale parameter $1/AT$, with its mode at $\hat{F} = n_{\rm obs}/(AT)$, which could serve as a convenient point estimator.
Expressed in terms of $\mu = ATF$, this gamma distribution has mean $\langle ATF\rangle = n_{\rm obs} + 1$ and variance $n_{\rm obs} + 1$.
For large $n_{\rm obs}$, the member likelihood function could be well approximated by a Gaussian function with mean and variance equal to $n_{\rm obs}+ 1 \approx n_{\rm obs}$:
\begin{align}\label{eq:GammaNormal}
h(F;n_{\rm obs}) 
  &\propto \frac{1}{n_{\rm obs}^{1/2}\sqrt{2\pi}}
     \exp\left[-\frac{(ATF - n_{\rm obs})^2}{2n_{\rm obs}}\right]\nonumber\\
  &\propto \exp\left[-\frac{(F - n_{\rm obs}/(AT))^2}
     {2n_{\rm obs}/(AT)^2}\right].
\end{align}
Thus for \emph{generating} simulated data, we would use a normal distribution, (\ref{eq:PoissonNormal}), with variance depending on the true flux.
But for \emph{analyzing} a simulated data set, we could use member likelihood functions either given by $g(n_{\rm obs};F)$ (Gaussian with respect to $n_{\rm obs}$, with variances depending on $F$, and thus \emph{non}-Gaussian with respect to $F$) or $h(F;n_{\rm obs})$ (Gaussian with respect to $F$, with variances depending on the simulated data, $n_{\rm obs}$, for each object).

For our simulations of catalog data, we use normal sampling distributions for estimated fluxes, $\Fhat_i$, with standard deviations that depend on the true fluxes, $F_i$, according to
\begin{align}
\sigma(F)
  &= \sqrt{\sigma_0^2+(\alpha F)^2}\nonumber\\
  &= \sigma_0 \left[1 + \left(\frac{\alpha F}{\sigma_0}\right)^2 \right]^{1/2},
\label{eq:sig-F}
\end{align}
where $\sigma_0$ characterizes the noise contributions from backgrounds and detector electronics, and $\alpha$ characterizes how Poisson fluctuations in the number of detected photons influence $\Fhat_i$, relative to the flux-independent noise sources.
For member likelihoods, we used the corresponding $g(n_{\rm obs};F)$ functions.
Based on published simulations of existing and anticipated surveys (\citealt{F99-SDSSSim,LSSTRefDesign}), we set $\sigma_0 = 6.4\times 10^{-10}$, $\alpha = 10^{-2}$, and \mbox{$\sigma_0 = 6.4\times 10^{-10}$~erg~cm$^{-2}$~s$^{-1}$} for our simulations.
The latter value was chosen so that sources with $L=20 u$ at $r=\rmax$ are just detectable by the error-based detection criterion described below.

% This was Tom's original choice; Janos used g(n;F) instead:
\iffalse
For an object with simulated best-fit flux $\Fhat_i$, we use a member likelihood function that is a Gaussian function with mode at $\Fhat_i$ and standard deviation parameter
\begin{equation}
\hat{\sigma}(\Fhat_i)
  = \sqrt{\sigma_0^2+(\alpha \Fhat_i)^2},
\label{eq:sig-Fhat}
\end{equation}
using the same values for $\sigma_0$ and $\alpha$ as are used for the sampling distributions.
\fi

% Checked by Janos 4/9/17:
% \enote{Check these error/threshold params; they should correspond to Janos's values shifted to physical units.}

For detection, we require a candidate object to have estimated flux a factor $\nu = 5$ times the measurement error.
This corresponds to a threshold flux satisfying $\Fth = \nu \hat{\sigma}(\Fth)$, which gives
\begin{equation}\label{eq:Fth}
\Fth = \frac{\nu \sigma_0}{\sqrt{1-\alpha^2}}.
\end{equation}
For the parameters of our simulation, this corresponds to \mbox{$\Fth \approx 3.2 \times 10^{-9}$~erg~cm$^{-2}$~s$^{-1}$}.
The detection efficiency is the probability that the measured value, $\Fth$, for a source with true flux $F$ will be above the threshold,
\begin{align}\label{key}
\eta(F)
  &= p(\Fhat > \Fth|F)\nonumber\\
  &= \Phi\left(\frac{F - \Fth}{\sigma(F)}\right),
\end{align}
where $\Phi(\cdot)$ denotes the standard normal cumulative distribution function.

\subsection{Case study}

We implemented this model in CUDAHM and used it to estimate population parameters and latent member parameters for a variety of simulated datasets based on the BB1 model.
Here we present results for an example with $100{,}000$ galaxies; below we examine performance vs.\ sample size.
The true values of the population parameters used for the simulation are listed in the first row of Table~\ref{tab:sum_est}.

%based on visual inspection of trace and autocorrelation plots for all population parameters, and a selection of latent member parameters.

\begin{figure}
    \centering
    \begin{subfigure}{0.3\textwidth}
        \includegraphics{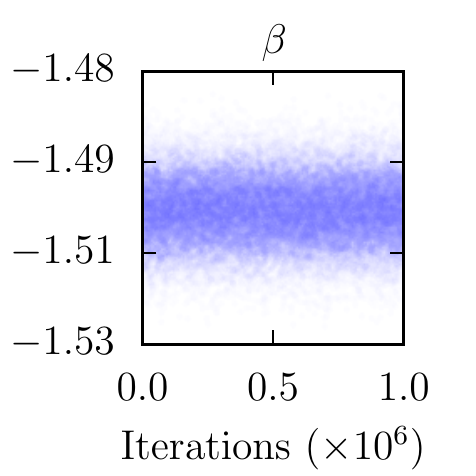}
    \end{subfigure}
    \begin{subfigure}{0.3\textwidth}
        \includegraphics{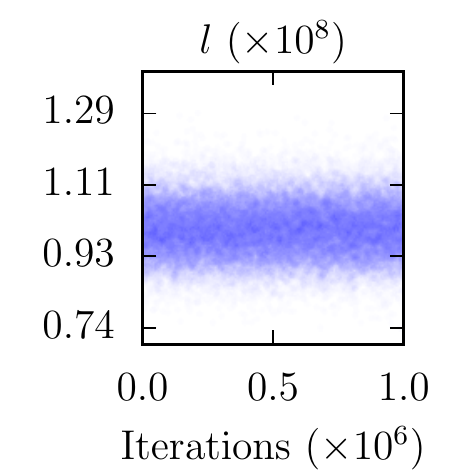}
    \end{subfigure}
    \begin{subfigure}{0.3\textwidth}
        \includegraphics{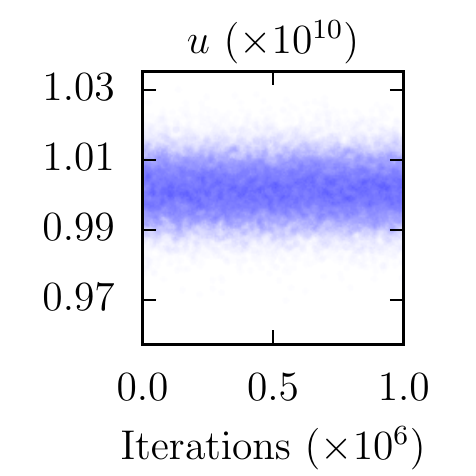}
    \end{subfigure}

% marginal posterior PDFs (middle row)
% The red vertical lines on the histograms show the true values of the parameters.
% Parameters $l$ and $u$ are in $L_\odot$ units.
%    \begin{subfigure}{0.3\textwidth}
%        \includegraphics{{fig/beta_histo}}
%    \end{subfigure}
%    \begin{subfigure}{0.3\textwidth}
%        \includegraphics{{fig/lowerscale_histo}}
%    \end{subfigure}
%    \begin{subfigure}{0.3\textwidth}
%        \includegraphics{{fig/upperscale_histo}}
%    \end{subfigure}    
    
    \begin{subfigure}{0.3\textwidth}
        \includegraphics{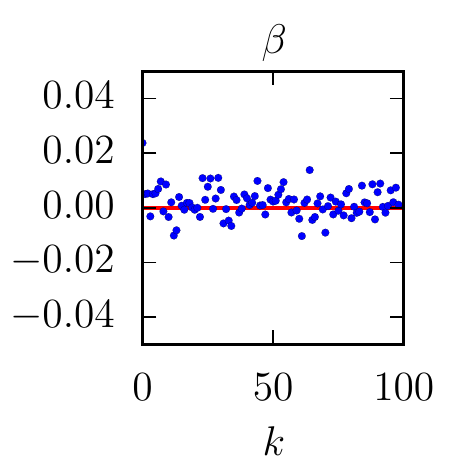}
    \end{subfigure}
    \begin{subfigure}{0.3\textwidth}
        \includegraphics{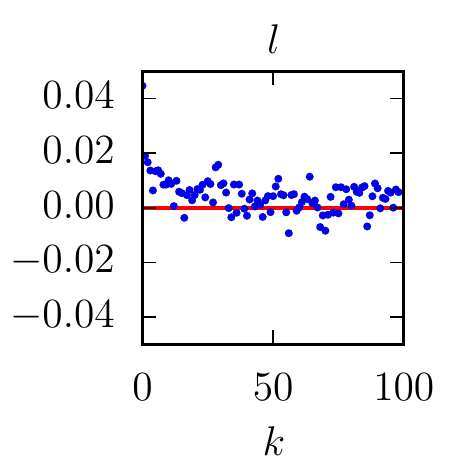}
    \end{subfigure}
    \begin{subfigure}{0.3\textwidth}
        \includegraphics{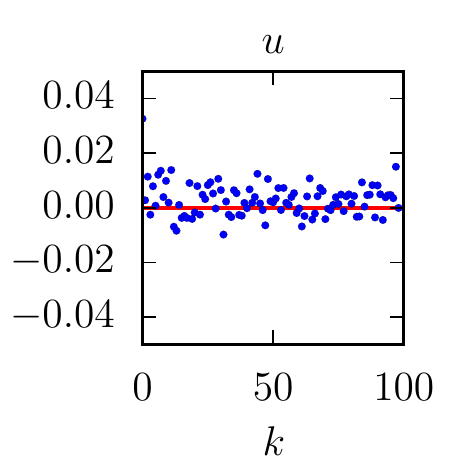}
    \end{subfigure}    
    \caption{Trace plots (upper row) and autocorrelation functions (lower row) based on thinned MCMC posterior samples of the three population parameters; $k$ is the lag for the autocorrelation (in units of the thinned-by-25 sample number).
    In the traceplots, samples are labeled by their unthinned step number.}
	\label{fig:results}
\end{figure}

\begin{figure}[tp]
\begin{center}
\includegraphics[width=.9\textwidth]{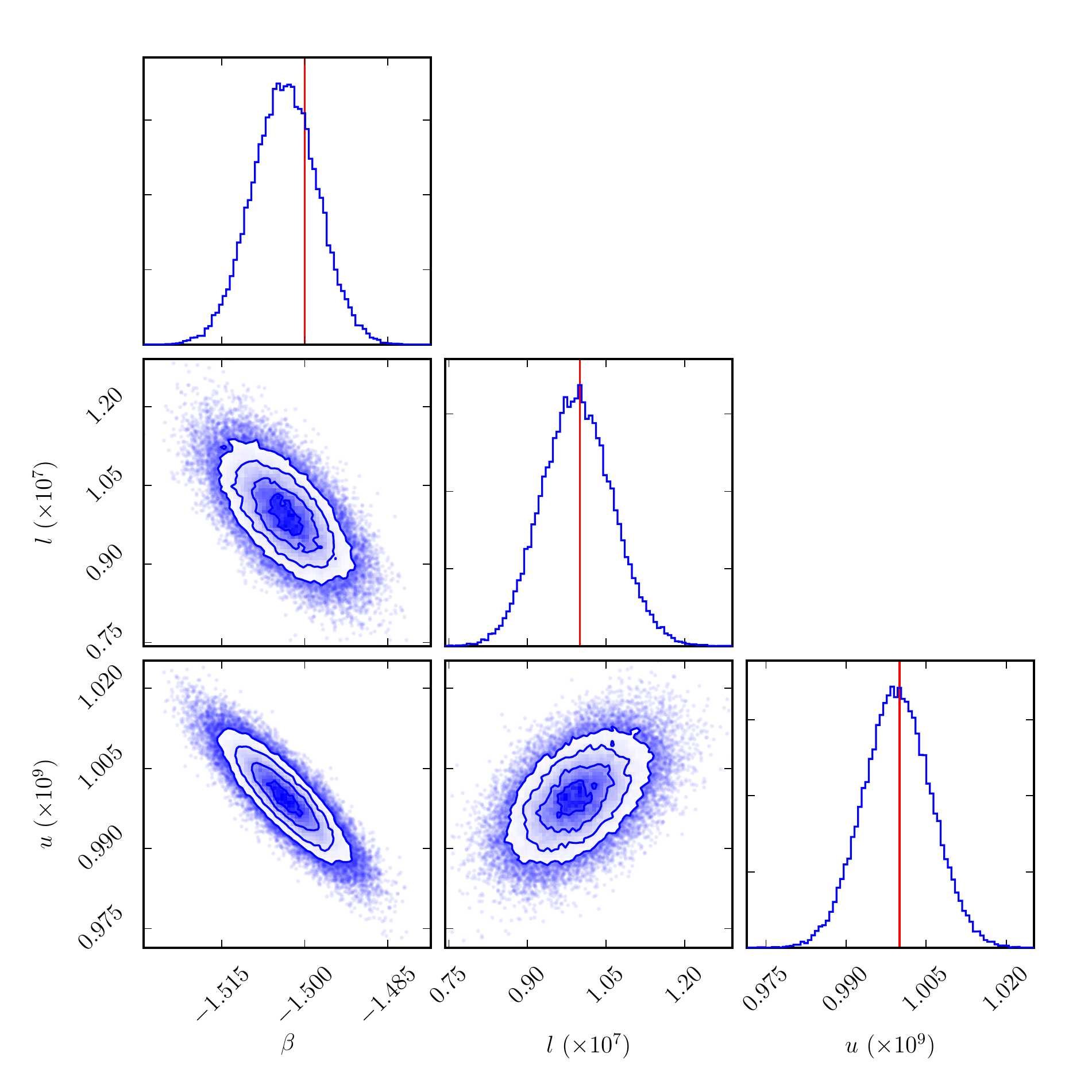}
\end{center}
\caption{Histograms and pairwise scatterplots depicting one- and two-dimensional marginal PDFs for the BB1 luminosity PDF parameters.
Vertical lines in histograms (red in the online version) indicate true values of parameters.}
\label{fig:pairs}
\end{figure}

We executed $2\times10^6$ burn-in steps and $1\times10^6$ post-burn-in MCMC steps to sample the probability distribution of the population parameters.
The length of the burn-in segment was chosen based on Geweke's heuristic test for failure of convergence, assessing consistency of means and variances of early and late segments of a potentially usable part of the full chain (\citealt{G92-MCMCConvergence}).
Based on the autocorrelation function of post-burn-in samples, the ESS for $\theta$ was $\approx 80,000$, much more than needed for standard inferences (e.g., computing credible regions).
We aggresively thinned (largely for convenience), keeping every $25^\textnormal{th}$ sample so the final number of samples was $40{,}000$.
Fig.~\ref{fig:results} shows trace plots and autocorrelation functions based on the population parameter samples.
The (thinned) chains appear to be well-mixed.
Table~\ref{tab:sum_est} lists point estimates (posterior means) and posterior standard deviations for the population parameters.
Fig.~\ref{fig:pairs} shows histograms and pairwise scatterplots depicting one- and two-dimensional marginal PDFs for the population parameters.
The parameters are accurately recovered, with high precision.

% This uses Janos's original dimensionless units:
%\begin{table} \begin{center} \begin{tabular}{ l | d{12} | d{12} | d{12} | } \cline{2-4} & \beta & l & u \\ \hline \multicolumn{1}{|l|}{$\theta_\text{true}$} & -1.5 & 5.0\hphantom{000} \times 10^{10} & 5.0\hphantom{000} \times 10^{12} \\ \hline \multicolumn{1}{|l|}{$\theta_{\text{MCMC}}$} & -1.5037 & 5.0302 \times 10^{10} & 5.0000 \times 10^{12} \\ \hline \multicolumn{1}{|l|}{$\sigma_{\theta\text{,MCMC}}$} & 0.0059 & 3.3548 \times 10^{9} & 3.1806 \times 10^{10} \\ \hline \multicolumn{1}{|l|}{$\theta_{\text{MLE}}$} & -1.5564 & 7.3222 \times 10^{10} & 5.7207 \times 10^{12} \\ \hline \multicolumn{1}{|l|}{$\theta_{\text{MLE, no noise}}$} & -1.5009 & 4.9341 \times 10^{10} & 4.9819 \times 10^{12} \\ \hline \multicolumn{1}{|l|}{$\lvert \theta_{\text{true}} - \theta_{\text{MLE}} \rvert$} & >9.5525 \cdot \sigma_{\beta\text{,MCMC}} & >6.9221\cdot\sigma_{l\text{,MCMC}} & >22.6591\cdot\sigma_{u\text{,MCMC}} \\ \hline \end{tabular} \end{center}\end{table}

\begin{table}
\begin{center}
\begin{tabular}{ l | d{12} | d{12} | d{12} | }
\cline{2-4}
  & \multicolumn{1}{|c|}{$\beta$} 
  & \multicolumn{1}{c|}{$l\; (L_\odot)$} 
  & \multicolumn{1}{c|}{$u\; (L_\odot)$} \\
\hline
 \multicolumn{1}{|l|}{$\theta_\text{true}$} & -1.5 & 1.0\hphantom{000} \times 10^{8} & 1.0\hphantom{000} \times 10^{10} \\
\hline
%  \multicolumn{1}{|l|}{$\hat\theta$} & -1.5036 & 0.99574 \times 10^{8} & 0.99962 \times 10^{10} \\
\multicolumn{1}{|l|}{$\hat\theta$} & -1.5036 & 0.996 \times 10^{8} & 0.9996 \times 10^{10} \\
\hline
%\multicolumn{1}{|l|}{$\sigma_{\theta}$} & 0.0059 & 0.068 \times 10^{8} & 0.00636 \times 10^{10} \\
\multicolumn{1}{|l|}{$\sigma_{\theta}$} & 0.0059 & 0.068 \times 10^{8} & 0.0063 \times 10^{10} \\
% \hline
%  \multicolumn{1}{|l|}{$\theta_{\text{MLE}}$} & -1.5564 & 1.46444 \times 10^{8} & 1.14414 \times 10^{10} \\
% \hline
%  \multicolumn{1}{|l|}{$\theta_{\text{MLE, no noise}}$} & -1.5009 & 0.98682 \times 10^{8} & 0.99638 \times 10^{10} \\
% \hline
%  \multicolumn{1}{|l|}{$\lvert \theta_{\text{true}} - \theta_{\text{MLE}} \rvert$} & >9.5226 \cdot \sigma_{\beta\text{,MCMC}} & >6.8309\cdot\sigma_{l\text{,MCMC}} & >606.4789\cdot\sigma_{u\text{,MCMC}} \\
\hline
\end{tabular}
\end{center}
\caption{Population parameters for simulation study with $10^5$ samples.
First row indicates the true values used to simulate data.
Second row lists posterior means; third row lists posterior standard deviations (both estimated using the MCMC output).
% For reference, we indicate the outcome of the ML estimatior run on the simulated data with and without noise in rows 4~and~5, respectively.
% To compare the Bayesian model to ML, the last row of the table shows the difference between the ML estimator and the true values in terms of the standard deviation of the posterior from the Bayesian model.
} \label{tab:sum_est} \end{table}

%................................................................................
\subsection{Performance test}

\begin{figure}
   	\begin{center}
   		\includegraphics{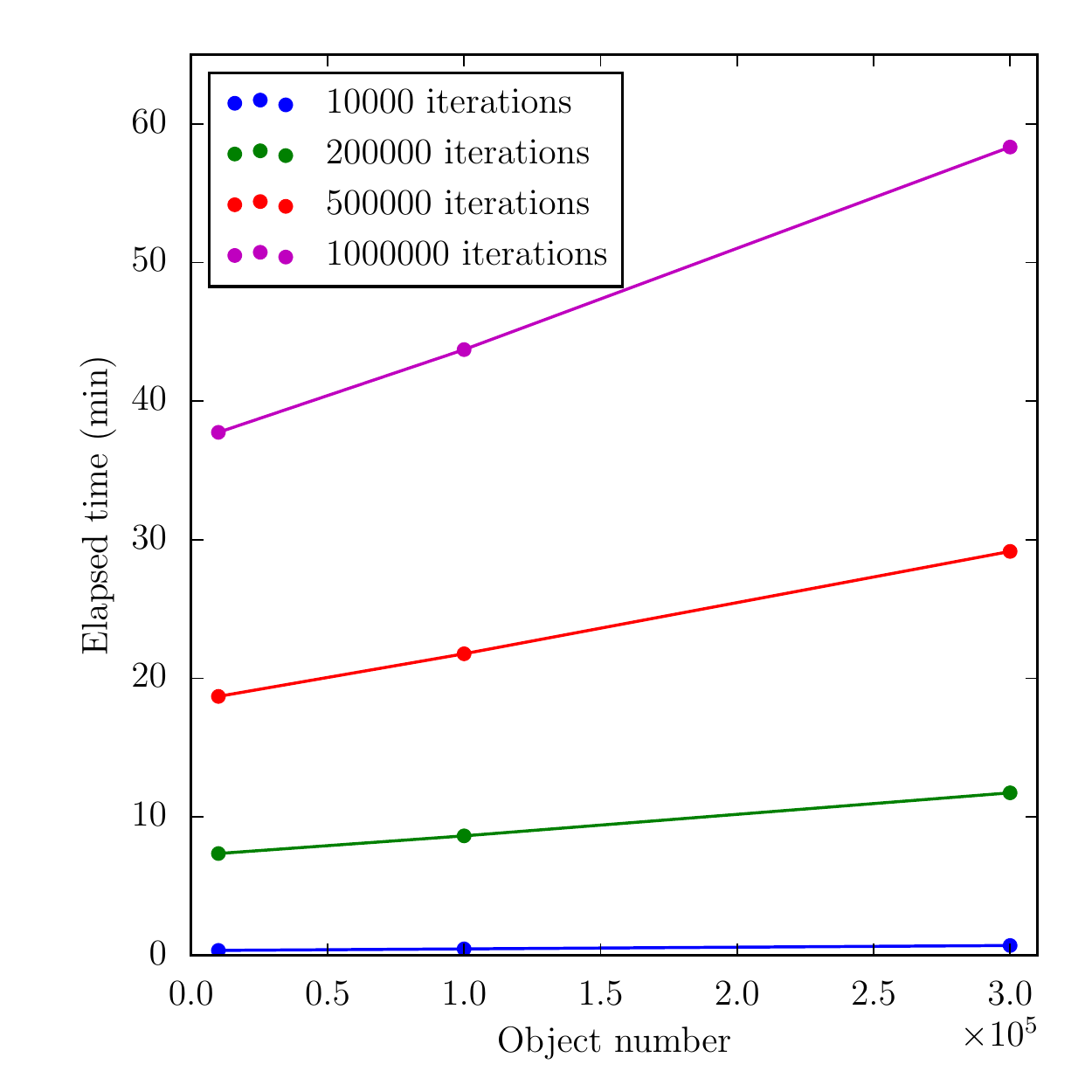}
   	\end{center} 
    \caption{Runtime of the CUDAHM code as a function of the number of objects, after a given number of iterations.
    The scaling of runtime with the number of objects is approximately linear.}
    \label{fig:performance_obj_vs_time}
\end{figure}

We used a single NVIDIA Tesla K40c card for performance tests; this card is optimized for scientific computation.
Fig.~\ref{fig:performance_obj_vs_time} shows the elapsed time as a function of the number of objects in the population, for various numbers of iteration steps (unthinned).
Inference for a population of $3\times10^5$ objects was achievable in one hour with a single Tesla GPU card.
The linear scaling of computation time with the number of objects (for fixed number of iterations, not fixed ESS) is due to the fact that the latent characteristics are conditionally independent, so that their sampling step in the MWG algorithm does not depend in a complicated way on the size of the latent parameter space.
More explicitly, the qualitative behavior displayed in Fig.~\ref{fig:performance_obj_vs_time} may be understood by considering the separate costs of the two steps of the MWG algorithm, i.e., sampling member properties, $\psivec$, from the population distribution (equation~(\ref{eq:psi_sampling})), and sampling new population parameters, $\ppar$, from their conditional distribution (equation~(\ref{eq:theta_sampling})).
Suppose the cost of population parameter sampling is $C_p$, and the per-member cost of latent property parameter sampling is $C_m$.
For each MWG cycle, there is one $\ppar$ sampling calculation, and $N$ independent $\psivec$ sampling calculations.
For $M$ MCMC steps, the total cost is
\begin{equation}
C = M C_p + M N C_m.
\label{cost}
\end{equation}
The upward march of the curves in Fig.~\ref{fig:performance_obj_vs_time} reflects the intercept (first term) in equation~(\ref{cost}), proportional to run length.
For a given curve, the growth with $N$ reflects the second term (presuming the thread pool is filled).
The slope of the curve grows with run length because of the $M$ dependence in this term.
To gain insight into the origin of the $C_p$ cost for this problem, we compared performance with a version of the problem without selection effects, eliminating the $\ppar$-dependent integral in the denominator of equation~(\ref{eq:epdf}), defining the effective density, $\epdf(F,r,\rhopar)$; this integral must be done by two-dimensional cubature for each new $\ppar$ proposal.
The comparison indicates this integral dominates $C_p$. 

% Real galaxy catalogs contain objects on the order of $10^8$, two magnitudes more than our simulated data set.
% Extrapolating from out performance numbers, estimating the parameters of the luminosity function with $2 \times 10^5$ Markov steps would take about 2000 minutes, a bit less then one and a half days, which makes applying our method to real data feasible.

Current astronomical catalogs for which the type of parametric modeling presented here is appropriate range in size from hundreds to well over $10^6$ objects.
Forthcoming catalogs will contain of order $10^8$ or more objects.
The example presented here shows that GPU-accelerated hierarchical Bayesian inference is feasible for such catalogs.
E.g., scaling from the BB1 example here, parametric inference with models with several parameters, for catalogs with $\sim 10^8$ objects, should be achievable with computing times of order a day with a modest-sized GPU clusters.

%===============================================================================
\section{Discussion}
\label{sec:discussion}

We have described a usefully general framework for massively parallel implementation of simple hierarchical Bayesian models using an adaptive Metropolis-within-Gibbs algorithm running on a GPU, exploiting conditional independence structure to accelerate computation.
The framework was motivated by population modeling problems in astronomy, where accounting for measurement error and selection effects is important, and straightforward to handle in a hierarchical Bayesian framework.
We addressed an example problem from this area---estimating galaxy luminosity functions.
This problem has some unique statistical features (in particular, the use of the same data for detection/selection, and for measurement).
We described a thinned latent marked point process framework for modeling such data.
Our implementation using CUDAHM demonstrates feasibility of hierarchical Bayesian modeling of existing and forthcoming large catalogs of properties of astronomical objects.

The CUDAHM implementation described here is the first step in an ongoing research program we are pursuing using GPUs to accelerate computations for hierarchical Bayesian models.
Our current work in progress is in two directions.

First, we are implementing GPU-accelerated cubature methods for inference with models that have low-dimensional latent characteristics per object.
In the example of the previous section, there was only a single uncertain latent parameter per galaxy, the flux, or equivalently, luminosity (because the redshift was considered to be precisely measured).
Similar univariate (conditional) density estimation problems arise for modeling cosmic populations of various types of objects---galaxies, active galaxies, stars, and minor planets.
Generalizing, many galaxy catalogs report imprecise redshifts; for such populations, the latent characteristics are two-dimensional---flux and redshift---or few-dimensional, when fluxes in multiple filters are available.
Objects observed in gamma rays or cosmic rays may have uncertain directions on the sky; this is a two-dimensional latent characteristic.
In contrast to much hierarchical modeling in other disciplines, astronomers are typically interested in estimating population parameters, and not in estimating object parameters.
In these settings, marginalizing over the latent characteristics could be done using quadrature and cubature algorithms rather than MCMC, producing a marginal likelihood function for the population parameters (which could be explored with a modest-dimensional MCMC algorithm).
Conditional independence makes this marginalization embarrassingly parallel.
In fact, the earliest work on hierarchical Bayesian demographic modeling in astronomy adopted this approach (e.g., \cite{LW98-GRBs-Iso,LW98-GRBs-Aniso}), as have some implementations of frequentist random effects models using maximum marginal likelihood methods (e.g., \cite{WT06-IRT-MaxMargLike}).
We are developing GPU implementations of cubature-based latent parameter marginalization for commonly arising scenarios, and studying how they compare with the fully MCMC-based approach described in this paper.

Our second ongoing research direction concerns a question of principle: with large populations, simple parametric population models, while potentially capturing the salient features of the population, are unlikely to faithfully describe detailed structure in the population distribution.
One possible response is to abandon parametric models completely, and explore nonparametric models.
However, parametric models are often motivated by important salient physics.
Astronomers thus may seek inferences that simultaneously recognize both the value of a parametric salient feature model, and the necessity to account for misfit in estimates of the parametric model.
This motivates a \emph{semiparametric} approach where a parametric model of interest is paired with a nonparametric discrepancy model.
Such semiparametric models have been developed in the context of regression based on data with additive noise (e.g., with Gaussian process discrepancy or residual models).
However, most cosmic demographic modeling is in the context of point process models.
For such cases, we are exploring semiparametric density deconvolution where the discrepancy is modeled via a \emph{multiplicative} component.
We are working on two appraoches.
The first multiplies the parametric model by an explicit sieve-like model, e.g., based on Bernstein polynomials, which are easily constrained to be non-negative.
The second multiplies the parametric model by an implicit gamma process; for a Poisson point process model, marginalization over the implicit discrepancy process produces a negative binomial marginal model.
GPU implementations should enable application of both approaches to large datasets.

% Mention multiplicative discrepancy factors?

%===============================================================================
\pagebreak[4]

\begin{center}
{\large\bf ACKNOWLEDGMENTS}
\end{center}

This material was based upon work partially supported by the National Science Foundation under Grant DMS-1127914 to the Statistical and Applied Mathematical Sciences Institute (SAMSI). 
The CUDAHM project began in a working group affiliated with the 2012--2013 SAMSI Program on Statistical and Computational Methodology for Massive Datasets.
This paper was completed as part of the 2016--2017 SAMSI Program on Statistical, Mathematical and Computational Methods for Astronomy.
Budav\'ari, Kelly, and Loredo are grateful to SAMSI for support of this work.
Loredo's work on this project was also supported by NSF grant AST-1312903.
Szalai-Gindl's work at Johns Hopkins University, where he was a visiting graduate student, was supported by the Hungarian Scientific Research Fund via grant OTKA~NN~114560.
Szalai-Gindl acknowledges useful discussions with his Ph.D.\ supervisors: Istv\'an Csabai and Laszl\'o Dobos.
Any opinions, findings, and conclusions or recommendations expressed in this material are those of the authors and do not necessarily reflect the views of the National Science Foundation.

%===============================================================================
\pagebreak[3]

\spacingset{1}

%================================================================================
\section*{Appendix A:\\Overview of CUDAHM API}
\renewcommand{\theequation}{A.\arabic{equation}}
\renewcommand{\thefigure}{A.\arabic{figure}}

% *** Put this in the README
% CUDAHM uses NVIDIA's \Cpp\ \texttt{Thrust} library for allocating storage on the host CPU and GPU.

CUDAHM enables \Cpp\ programmers to rapidly construct a MWG sampler for a simple hierarchical model, requiring the user to supply only a minimimal amount of CUDA code.
A template file in the distribution, \texttt{cudahm\_blueprint.cu}, contains a heavily commented example application that users may copy and customize.
Here we briefly highlight key elements of the application programming interface (API).

For basic applications, the user must define two functions, instantiate a single class that manages the computations, and create a configuration file read by the compiled application that specifies parameters defining an MCMC run.
The functions compute probabilities needed by the two steps in the MWG algorithm.
One computes the logarithm of the probability density for the measurements given the member properties for each object in the sample (the product of the $\mlike(\cdot)$ member likelihood functions).
The second computes the logarithm of the probability density for the member properties given the parent population parameters (the product of $f(\cdot)$ population distribution factors).
Both functions execute on the GPU and must be written as CUDA kernels (i.e., using the CUDA extensions to the \texttt{C} language).

To manage the computations, the user must instantiate one of two classes.
A \texttt{GibbsSampler} instance runs the MWG algorithm, storing all samples in memory; the user can write out the samples, or a subset, as needed upon completion.
Alternatively, a \texttt{Gibbs\-Sampler\-With\-Compact\-Memory\-Usage} instance runs the algorithm, but with more efficient use of memory; it opens an output file stream (which has a buffer) and writes samples out on the fly.

Internally, these sampler classes instantiate two classes that users may wish to customize (by subclassing) for advanced use cases (in which case pointers to the subclasses must be provided to the sampler's constructor).
The \texttt{DataAugmentation} class controls calculations involving the member properties.
The \texttt{PopulationPar} class controls calculations involving the population parameters.
Users may wish to subclass these classes in order to customize one or both steps of the MWG algorithm.
For example, evaluation of the prior for population parameters occurs in a \texttt{PopulationPar} instance, which implements a uniform prior as a default choice.
To override this default prior, the \texttt{PopulationPar} class should be subclassed.

We note the CUDAHM default methods assume \emph{all} member properties are uncertain; kernel functions that are used for updating member properties and the population parameters on the GPU assume all member properties may change in each MWG cycle.
This is important to consider for problems where the object properties contain precisely measurable predictors (e.g., covariates for conditional density estimation in the manner of the middle DAG in Fig.~2.2), which should be held fixed over the course of posterior simulation.
When this is the case, the user should override the default methods.
The \texttt{lum\_func} implementation, which is related to the luminosity function example case, provides concrete guidance on this issue.
We are exploring internal architectural changes to simplify the user API for handling such cases.

% TODO: Provide a license for CUDAHM.

% TODO: Add a Makefile and README to the repo!

CUDAHM is open-source software maintained on GitHub at \url{https://github.com/tloredo/CUDAHM}; the branch named \texttt{paper1} contains the version described in this paper and can be obtained via the following Git command:
\begin{verbatim}
git clone -b paper1 https://github.com/tloredo/CUDAHM [directory]
\end{verbatim}
where \verb|[directory]| denotes an optional name for the destination directory (overriding the default \verb|CUDAHM|).
The CUDAHM distribution contains example code implementing basic hierarchical models, such as the normal-normal model, and a realistically complicated astrophysical example handling regression with classical measurement error, with nonlinear models (the interstellar dust problem briefly described in the main text at the end of \S~2.1).

%================================================================================
%\section*{Appendix B: Smoothly truncated \\break-by-one power law luminosity function}
\section*{Appendix B: Exponential-cutoff \\break-by-one power law luminosity function}
\renewcommand{\theequation}{C.\arabic{equation}}
\renewcommand{\thefigure}{C.\arabic{figure}}

This Appendix describes the \emph{break-by-one gamma distribution} we use as a luminosity function in our example calculations.
For further discussion of this luminosity function model, see \cite{L20-BB1Gamma}.
 
For most cosmic populations, including galaxies, the luminosity function falls very steeply with increasing luminosity.
The canonical starting point for parametric modeling of luminosity distributions is the \emph{Schecter function}, a power law that is smoothly truncated at large luminosities by an exponential decay factor:
\begin{equation}
\lfunc(L;\theta) =
  \frac{A}{L_*} \left(\frac{L}{L_*}\right)^{\beta} e^{-L/L_*},
\label{eq:schecter}
\end{equation}
where the parameters $\theta = (L_*,\beta, A)$ comprise a luminosity scale, $L_*$, a nominal mid-luminosity power law index, $\beta$, and an amplitude, $A$.
(There are varying conventions for parameterizing the amplitude of the Schecter function.
In this parameterization, $A$ has units of space density.
In similar parameterizations, $A$ is often denoted $\lfunc_*$, although it neither has the units of $\lfunc$, nor is it equal to $\lfunc(L_*)$, as the symbol might misleadingly suggest.)
% Astronomers very typically fit such simple distributions to LF data.
The Schechter function follows from a basic physical model of galaxy formation via self-similar gravitational condensation (\citealt{PS74-SelfSimSchechter}).
Many astrophysical processes are approximately self-similar over a wide range of scales, leading to power law distributions.
The index provides information about the scale-free behavior; upper and/or lower breaks provide information about key physical scales.
Astronomers often seek to distill these kinds of salient features from the data.

The form of the Schecter function would seem to imply a luminosity distribution that is a gamma distribution (with shape parameter $\alpha = \beta - 1$).
However, the observed samples of many populations follow Eq.~\ref{eq:schecter} with $\beta$ in the interval $(-2,-1)$, in which case the integral of $\lfunc(L;\theta)$ over $L$ is infinite, and the luminosity distribution is formally improper (with $\alpha$ outside of the allowed range for the gamma distribution).
Low-luminosity sources are unobservable (due to noise and background, discussed in the main text), so in practice the \emph{observable} luminosity function is truncated at low luminosities, and the impropriety is often ignored.
But the actual luminosity function must rise less quickly with decreasing $L$ (corresponding to $\beta$ becoming larger than $-1$) or be cut off at low luminosities (corresponding to there being a minimum galaxy size).

For some populations, an increase in the power law index (i.e., flattening of the logarithmic slope) is in fact observed at low observable luminosities.
For example, this is the case for quasars (galaxies with a large, actively accreting central black hole; see \citealt{M+13-QuasarLumFunc}).
Similarly, the stellar initial mass function (related to the stellar luminosity function, and fit with similar models) has a low-mass (low-luminosity) index that flattens by $\approx 1$ (\citealt{K07-IMF-BPL}).
Motivated by such observations, and to keep the luminosity distribution proper, we here adopt a ``break-by-one'' (BB1) generalization of the Schecter function, with $\lfunc \propto L^{\beta+1}$ at low luminosities, and thus integrable for $\beta > -2$.

Specifically, the BB1 model has a luminosity distribution with three parameters: a mid-luminosity power law index, $\beta$, and two parameters defining the mid-luminosity range, $(l, u)$, with $l < u$ and  $u$ playing the role of $L_*$ in the Schecter function, and the power law index smoothly breaking to $\beta+1$ as $L$ decreases below $l$.
The BB1 luminosity PDF has following functional form:
\begin{equation}
\label{aeq:lumPDF} 
\lpdf(L ; \theta) = 
  \frac{C(\beta,u,l)}{u}\left(1-e^{-L/l}\right) \left(\frac{L}{u}\right)^{\beta} e^{-L/u},
\end{equation}
where the normalization constant $C(\beta,u,l)$ is
\begin{equation}
\label{eq:normLumPDF} 
C(\beta,u,l) =
  \begin{cases} \dfrac{1}{\Gamma(\beta+1)\cdot\left(1-\frac{1}{\left(1+\frac{u}{l}\right)^{\beta+1}}\right)} 
    & \quad \text{if } \beta > -2\text{ and }\beta \ne -1; \\
 \dfrac{1}{\log\left(1+\frac{u}{l}\right)} & \quad \text{if } \beta=-1.
  \end{cases}
\end{equation} 
Note that as $l\rightarrow 0$, the BB1 distribution becomes a gamma distribution (if $\beta > -1$).
%Also, in our computational implementation, the condition $\beta=-1$ of the first case is $-1.001<\beta<-0.999$.
We designed the BB1 distribution to have smooth power law break behavior at low $L$, yet also have an analytical normalization constant;
it is proper for $\beta > -2$.
We generate samples from the BB1 distribution using a straightforward modification of a widely-used algorithm for sampling from the gamma distribution (\citealt{ahrens_computer_1974}).
These properties make it useful for simulation experiments.

We define a BB1 luminosity function by multiplying the BB1 luminosity distribution by the galaxy spatial number density, which is simply a constant, $n$, for a homogeneous population.

%  Fig.~\ref{fig:lumfunc}

Fig.~4.1 (in the main text) shows an example BB1 luminosity function, with $\beta = 1.5$, and $(l,u) = (1\times 10^{8}, 1\times 10^{10})$ in solar luminosity ($L_\odot$) units; it is plotted both with $\log$-linear axes, and with $\log$-$\log$ axes, where the varying power law behavior is evident.
The local power law index corresponds to the slope, $G(L)$, in $\log$-$\log$ space, defined by
\begin{equation}
	G(L) \equiv \frac{\dd\log{\lpdf}}{\dd\log{L}} = \frac{L}{\lpdf} \frac{\dd \lpdf}{\dd L} = g(L) + \beta - \frac{L}{u},
\end{equation}
with
\begin{equation}
	g(L) = \frac{L}{l}\cdot\frac{1}{e^{L/l} - 1}.
\end{equation}
Evidently, $g(L) \rightarrow 0$ for $L \gg l$ and $g(L) \rightarrow 1$ for $L \ll l$.
Thus the logarithmic slope, $G(L)$, corresponds to an exponential cutoff at large $L$, and at small $L$, a slope of $\beta + 1$.
When $u\gg l$, so there is a range where $L\gg l$ but $L\ll u$, the logarithmic slope is $\approx \beta$ in that range.

Finally, the BB1 cumulative distribution function is 
\begin{equation}
\label{eq:lumCDF} 
\lcdf(L ; \theta) = 
  C(\theta)
  \left[ \Gamma(\beta + 1) - \gamma\left(\beta + 1, L/u\right) - \frac{\Gamma(\beta + 1)-\gamma\left(\beta + 1, L\cdot\left(\frac{1}{u}+\frac{1}{l}\right)\right)}{\left(1 + \frac{u}{l}\right)^{\beta + 1}} \right],
\end{equation}
where $\Gamma(\cdot)$ and $\gamma(\cdot, \cdot)$ denote the gamma function and the upper incomplete gamma function, respectively.

%===============================================================================
\bibliographystyle{agsm}

\bibliography{CUDAHM}

\end{document}